\documentclass[twocolumn]{aastex63} 

\usepackage{float}
\usepackage{graphicx}           
\usepackage{latexsym}           
\usepackage{amsmath}            
\usepackage{nccmath}            
\usepackage{bm}                 
\usepackage{color}              
\usepackage{booktabs}           
\usepackage{multirow}           
\usepackage{natbib}

\newcommand\n            {\noindent}

\newcommand\m            {\medskip}
\newcommand\s            {\smallskip}

\newcommand\bn           {\bigskip\noindent}

\newcommand\cl           {\centerline}

\newcommand\eg           {{\it e.g.},}
\newcommand\ie           {{\it i.e.},}
\newcommand\EBminV       {{$E_{\rm B-V}$} }
\newcommand\fesc         {{$f_\mathrm{esc}$}} 
\newcommand{\HST}        {\emph{HST}}
\newcommand\Lya          {{Ly$\alpha$} }

\newcommand\mum          {{$\mu$m}}
\newcommand\muJy         {{$\mu$Jy} }

\newcommand\degree       {{\ifmmode^\circ\else$^\circ$\fi}}  
\newcommand\arcm         {{\ifmmode {'\ }\else$'     $\fi} } 
\newcommand\arcs         {{\ifmmode{''\ }\else$''    $\fi} } 

\newcommand\arcspt       {{$\buildrel{\prime\prime}\over .$}}

\newlength{\txw}
\newlength{\txh}

\newcommand{\del}[1]{\relax}%
\newcommand{\DELETED}[1]{\relax}%
{\relax}%

\defcitealias{Jiang2020}{J20}

\shorttitle{Jeon, J., et al.}
\shortauthors{Jeon, J., et al.}

\begin{document}

\title{SED Analysis of 13 Spectroscopically Confirmed Galaxies at z$\simeq$6
to Constrain UV-Slope, Model Dust Attenuation and Escape Fractions}

\correspondingauthor{Junehyoung Jeon}
\email{junehyoungjeon@utexas.edu}

\author[0000-0002-6038-5016]{Junehyoung Jeon}
\affiliation{Department of Astronomy, University of Texas, Austin, TX 78712, USA}
\author[0000-0001-8156-6281]{Rogier A. Windhorst}
\affiliation{School of Earth \& Space Exploration, Arizona State University, Tempe, AZ 85287-1404, USA}
\author[0000-0003-3329-1337]{Seth H. Cohen}
\affiliation{School of Earth \& Space Exploration, Arizona State University, Tempe, AZ 85287-1404, USA}
\author[0000-0003-1268-5230]{Rolf A. Jansen}
\affiliation{School of Earth \& Space Exploration, Arizona State University, Tempe, AZ 85287-1404, USA}
\author[0000-0002-0648-1699]{Brent M. Smith}
\affiliation{School of Earth \& Space Exploration, Arizona State University, Tempe, AZ 85287-1404, USA}
\author[0000-0001-6650-2853]{Timothy Carleton}
\affiliation{School of Earth \& Space Exploration, Arizona State University, Tempe, AZ 85287-1404, USA}
\author[0000-0003-1344-9475]{Eiichi Egami}
\affiliation{Steward Observatory, University of Arizona, 933 North Cherry Avenue, Tucson, AZ 85721, USA}
\author[0000-0002-0496-1656]{Kristian Finlator}
\affiliation{New Mexico State University, Las Cruces, NM 88003, USA}
\author[0000-0003-4176-6486]{Linhua Jiang}
\affiliation{Kavli Institute for Astronomy and Astrophysics, Peking University, Beijing 100871, China}
\author[0000-0001-9298-3523]{Kartheik G. Iyer}
\affiliation{Dunlap Institute, University of Toronto, Toronto, ON, Canada}
\author[0000-0001-7166-6035]{Vihang Mehta}
\affiliation{Minnesota Institute for Astrophysics, University of Minnesota, 116 Church Street SE, Minneapolis, MN 55455, USA}

\begin{abstract}
The reionization of the hydrogen in the Universe is thought to have completed
by redshift $z\simeq5.5-6$. To probe this era, galaxy observations in the Subaru
Deep Field (SDF) have identified more than 100 galaxies at $z\simeq6$, many
spectroscopically confirmed through follow-up observations. We model the
spectral energy distributions (SEDs) of 13 SDF galaxies with the CIGALE and
Dense Basis codes using available optical/IR data. 
Modeling deep IR photometry has the potential to constrain the galaxy's Lyman
continuum (LyC) escape fraction (\fesc). We use the modeled nebular
emission lines and find that the implied escape fractions ranges from \textbf{0 to 0.8
with a median of $\sim$0.35 for Dense Basis and $\sim$0.55 for CIGALE.}
Significant uncertainties in the data exist, so that fitting results in a large
range of \fesc\ for individual objects. The implied median \fesc-values may be
high enough for galaxies to finish reionization by $z\sim6$. Furthermore, we
find no strong trends between the UV-slope $\beta$ or \EBminV with model \fesc.
If true, the lack of trends suggest that other factors besides nebular
emission or dust extinction could have led to LyC escaping, such as the
presence of holes in the ISM with sufficiently wide opening angles from
outflows of supernovae and/or weak AGN, resulting in a range of implied
\fesc-values depending on the viewing angle of each galaxy. The current
\textit{HST, Spitzer} and ground-based photometric and model errors for the
galaxies remain large, so IR spectroscopic observations with the \textit{James
Webb Space Telescope} are needed to constrain this possibility.
\end{abstract}


\bn \section{Introduction} \label{sec:intro}

Detailed studies of high redshift galaxies observed in the first billion years
have provided a wealth of information about early galaxy formation and
evolution. Developments in ground- and space-based telescopes, such as the
availability of very wide-field cameras on 8 to 10m class telescopes and
medium- or narrow-band filters that fall between the worst sky lines, enabled
the identification of numerous high redshift galaxies (\eg\
\citealt{Ellis2013,Jiang2013,Bouwens2015,Ota2017, Finkelstein2019,Hu2019}). In particular, observations with the
Subaru telescope, \textit{HST}, and the \textit{Spitzer Space Telescope} have
discovered numerous high redshift galaxies in optical, near-IR, and mid-IR,
respectively (\eg\ \citealt{Yan2010,Lorenzoni2011,McLure2011, Kashikawa2011, Trenti2011}). Large
areas of sky have been imaged from the ground with these narrow-band filters,
and the detected narrow-band emission of some of the brighter galaxies have
been spectroscopically confirmed to be Lyman $\alpha$ (Ly$\alpha$) emission
(\eg\ \citealt{Finkelstein2013,Jiang2013,Konno2014, Jiang2016,Laporte2017,stark2017,Ouchi2018}). The \textit{James Webb Space
Telescope (JWST)}, which can access wavelengths presently inaccessible with
\textit{Hubble Space Telescope (HST)} and that are difficult to access from
the ground due to Earth's atmosphere, will be able to observe these objects at
higher resolution to identify AGN, star-formation regions, and outflows (\eg\
\citealt{Gardner2006, Maseda2019, Windhorst2018}). As reionization of the
intergalactic hydrogen in the Universe is thought to have been completed by 
$z\simeq5.5-6$ \citep[\eg\ ][]{Keating2020}, such high redshift galaxies can
help us study this interesting period.

Some parameters of these high redshift galaxies useful for studying the epoch of
reionization include the UV-continuum slope ($\beta$), which can help constrain
the characteristics of their young stellar populations (such as age
\citealt{Conroy2013,Jiang2020}), their Lyman Continuum (LyC) escape fraction (\fesc) which
provides a measure of how much energy was available for reionization
\citep[\eg\ ][]{Miralda2000,Loeb2001,barrow2020}, and the \EBminV\ reddening
values \citep[\eg\ ][]{Calzetti2000, Meurer1999}, which are a measure of their
internal attenuation by dust. \EBminV or $\beta$ may be correlated with \fesc,
given that $\beta$ is related to the amount of ionizing emission originating
from the galaxy, and \EBminV is related to the amount of dust and gas preventing
that emission from escaping the galaxy \citep[\eg\ ][]{Ono2010, Hayes2011}. 

Previous studies have suggested that a minimum escape fraction of $\sim$20$\%$
is needed to finish the reionization of the hydrogen in the Universe at
$z\simeq6$, assuming UV-bright sources ($M_{AB}\lesssim-21$ mag) dominate
reionization \citep[\eg\ ][]{Finkelstein2012,Robertson2015, Naidu2020}. However, it is
unclear what \fesc-values are typical for $z\simeq6$ galaxies. For
galaxies at $z<4$, escape fraction may be directly measured through 
rest-frame UV imaging and/or spectroscopy. For galaxies at high redshift like
those in our sample, such direct methods do not work, since a larger portion of
LyC photons are absorbed by the higher HI fraction in the IGM at $z>4$, and any
constraints must rely on the more limited IR data that are available for our 
galaxies \citep{bian2020}.

Hence, we turned to an alternative method of SED fitting to infer the
escape fraction of our high redshift galaxies. Previous studies have noted that
modeling nebular emission is crucial in producing accurate SED models,
particularly at high $z$ \citep[\eg\ ][]{schaerer2009,Jiang2016}. This is
related to \fesc , since significant nebular emission is indicative of ionizing
photons exciting the gas to produce rest-frame UV and optical emission lines
instead of escaping the galaxy, implying lower \fesc values. Thus, SED modeling
can be used to {\it indirectly} constrain the fraction of ionizing photons that
is not absorbed and reprocessed as emission lines, but instead escapes from
galaxies.

Moreover, several authors found unusually blue $\beta$ values as steep as 
$\beta\simeq$ $-2\sim-3$ for faint galaxies at $z\geq6$ using photometric
methods \citep[\eg\ ][]{bouwens2014, Labbe2010, Jiang2020}. To see how such
surprisingly blue slope parameters might occur, SED modeling --- which
previously has been used to characterize galaxy populations at low
\citep{papovich2001, Shapley2001, Shapley2005} and moderate redshifts
\citep{Ono2010} --- can be used to interpret the UV-slope $\beta$. 

To perform this modeling on a large number of galaxies in the Subaru Deep
Field (SDF), 67 candidate galaxies at $z\simeq6$ were identified in
\citet{Jiang2013}\footnote{Five of the spectroscopically confirmed galaxies are
not in the SDF, but rather the UKIDSS Ultra-Deep Survey (UDS), with WFC3-IR
coverage from the Cosmic Assembly Near-infrared Deep Extragalactic Legacy
Survey (CANDELS; \citealt{grogin2011, koekemoer2011})} to be the most luminous
galaxies in \Lya emission and/or in UV-continuum in this redshift range around
$z\simeq6$. Of these galaxies, 7 were chosen to have particularly blue
$\beta$ and were imaged by \textit{HST} Wide Field Camera 3 (WFC3) in the
F105W, F110W, F125W, and F160W filters, and their rest-frame UV SEDs
were modeled with the CIGALE program \citep{Boquien2019}, as described in
detail in \citet[][hereafter \citetalias{Jiang2020}]{Jiang2020}.
\del{
Of these 7 galaxies, 6 {\bf showed observed \textit{HST} photometry that
confirmed} extremely blue $\beta$-values ($\beta\lesssim$ --2.5), which presents
a challenge to stellar population synthesis models, as the galaxies can yield
such blue slopes only under the assumption of extremely young ages and low
metallicities. We refer to these seven galaxies of \citetalias{Jiang2020} as
the {\bf ``7 galaxies of \citetalias{Jiang2020}'' throughout the text}. 
}

In this paper, we study the SEDs of \textbf{13 galaxies with available near-IR data, including 1 galaxy from \citetalias{Jiang2020}}, and compare these samples. Our goal for this comparison is to
determine if, and to what extent, galaxy parameters are correlated with LyC
escape fraction, and how galaxy properties are related to the slope of the UV
continuum. We use the CIGALE code for our SED models, which was also used by
\citetalias{Jiang2020} for their sample of galaxies. In addition, we performed SED
fitting using the Dense Basis code \citep{Iyer2017} to ensure that our results
were not sensitive to the particular SED-fitting code applied. For all 13 SDF
galaxies, we not only fit the stellar age, but also explore the values of the
LyC escape fraction (\fesc) and dust extinction (\EBminV{}) allowed for each
galaxy to determine their best SED fit. This paper thus presents the SED models
of the \textbf{13} $z\simeq6$ galaxies in the SDF that have near-IR data to possibly
constrain their escape fraction. We further
discuss the range of SED parameter values permitted by the CIGALE and Dense
Basis models.

The paper is organized as follows. In section \ref{2}, we describe all
available data for these $z\simeq6$ galaxies, as well as archival UKIRT/WFCAM
K-band data. In section \ref{3}, we describe how the SED modeling of the
galaxies was performed, and describe the results. In section \ref{5}, we discuss constraints to the
implied \fesc, $\beta$, and \EBminV values for the sample of 13 SDF
galaxies at $z\simeq6$ and investigate possible trends. We then summarize our
results in section \ref{6}. All AB magnitudes from \citetalias{Jiang2020} were
converted to mJy to facilitate direct modeling with the CIGALE and Dense Basis
programs \citep{Oke1983}. They are presented as \muJy in this paper.

\section{Overview of Extant Galaxy Data Used} \label{2}

Table~\ref{table1} lists IDs, coordinates, redshifts, and measured fluxes of the
13 galaxies in our sample as given in \citet{Jiang2013,Jiang2016} and
\citetalias{Jiang2020}. {\bf We also include the UV-slope measurements for the
the galaxy from \citetalias{Jiang2020}, ID43}. \HST/WFC3 images in the F105W, F110W, F125W, and/or F160W
filters are available in various combinations for all {\bf 13} galaxies, and our sample also has \textit{Spitzer}/IRAC imaging available in
the SDF, or from the larger area covered by the Subaru XMM-Newton Deep Survey
\citep{Furusawa2008} from \citet{Jiang2013,Jiang2016}. Sub-pixel
dithering was performed to improve the Point Spread Function (PSF) sampling
\citep{Jiang2013}. For the optical \HST\ images, photometry was performed using
SExtractor \citep{Bertin1996} adopting a 2\arcs diameter circular aperture and
subtracting the local background. We applied an aperture correction based on
bright point sources within the same image to account for any light losses
outside each object aperture. Rather than homogenizing PSFs to a given value,
PSF matching was done by multiplying the weight maps by the inverse square of
the PSF Full Width at Half Maximum (FWHM). For the infrared images, we used
SExtractor with a Kron factor of 1.8 to determine the total magnitudes, and the
same aperture corrections were applied.

\textbf{For the IRAC 3.6 and 4.5 \mum\ data, \citet{Jiang2016} deblended sources first by modeling the
brightest neighbors of the source with iGALFIT \citep{Ryan2011}, convolving
with the PSF image, and then subtracting the neighbors from the image.}
Photometry was then done on a 1\arcspt 8 radius aperture with an aperture
correction of 0.4 mag \citep{Jiang2016}. 

Since CIGALE provides SED models into the observed infrared regime, we
searched the \textit{Herschel} archive for images of the galaxies to possibly
constrain their dust content. In the region of the SDF containing our \textbf{13}
galaxies, extant \textit{Herschel} far-IR observations are too shallow for
robust detections of any of our sample galaxies. Hence, we have no far-IR
data to place further constraints on thermal dust emission in our CIGALE
modeling.

\del{Reversed order of next two paragraphs for better flow: }

To obtain the K-band data for the galaxies of \citetalias{Jiang2020}, we
searched the WFCAM Science Archive\footnote{http://wsa.roe.ac.uk/}. After
identifying the exposures containing each of the galaxies, we used the DS9
program \citep{Joye2003} to analyze the K-band image files  from LAS or
UDS. We selected a circular region around the location of each galaxy and a
circular region of background sky devoid of object signal, their sizes based on
the resolution of each image, to estimate the source and background flux
measurements on the Vega magnitude system, given the zero-point of
\citet{Hodgkin2009}. Using the region analysis function of the DS9
program, the sum of the pixel values in the regions were used to measure the
flux, and the standard deviation of the background pixel values was used to
estimate the uncertainties of the flux values. For the K-band, we used the
conversion from Vega to AB magnitudes from \citet{Hewett2006}. We converted the
apparent AB-magnitude values to $\mu$Jy units \citep{Oke1983}. 

Deep observations as part of the UKIRT WFCAM Large Area Survey (LAS) or the
Ultra Deep Survey (UDS) presented in \citet{Lawrence2007} observed the
SDF galaxies of \citetalias{Jiang2020} in the K-band, and provided 2.2 \mum\
flux measurement or upper limits. In those cases, the galaxies have 5--6
independent flux measurements to constrain the range of implied physical
parameters allowed by the CIGALE models. \textbf{Out of the 67 SDF galaxies including the 7
analyzed by \citetalias{Jiang2020}}, 14 galaxies did not have any reliable flux
measurements. An additional 6 galaxies only had one reliable flux measurement,
so they were also omitted. We also omitted 4 galaxies that had
neither K-band nor IRAC measurements --- \ie\ no measurements beyond 2 \micron
--- which did not allow meaningful modeling of their escape fractions. In the
end, the K-band data was not used for these galaxies, since it did not improve
the modeling of the 7 galaxies significantly. \textbf{Finally, we rejected 2 galaxies with signal to noise greater than 3 in the IRAC1 data, 15 galaxies with only an lower limit on the \textit{Spitzer} data, 9 galaxies with nearby neighbors within 1.7-2.0'' IRAC beam in the HST images, and 4 galaxies that did not readily have images that we could check for neighbors. Thus, 13 galaxies could be
modeled by CIGALE out of the full sample of 67 SDF galaxies. The visual cutouts of the galaxies can be found in \citet{Jiang2013}. }

\textbf{ID43 was} observed in the UKIRT WFCAM LAS with shallow observations. A 2$\sigma$ upper limit to the K-band fluxes is therefore listed for ID43 in Table~\ref{table1}.

\clearpage

\movetabledown=5.5cm
\begin{rotatetable}
\begin{deluxetable*}{cccccccccccccc}
\tabletypesize{\scriptsize}
\setlength{\tabcolsep}{2pt}
\tablenum{1}
\tablecolumns{13}
\tablecaption{\textbf{High redshift galaxy flux data at various \textit{HST} and \textit{Spitzer} filters with measured UV-slope from \citetalias{Jiang2020}}\label{table1}}
\tablewidth{0pt}
\tablehead{
\colhead{ID} &\colhead{R.A.} &\colhead{Decl.} & \colhead{z} & 
\colhead{Slope} & \colhead{$z'$} & \colhead{$y'$} &
\colhead{F105W}&\colhead{F110W} &\colhead{F125W} &\colhead{F160W} & \colhead{IRAC 1} &
\colhead{IRAC 2} & \colhead{K}\\
\colhead{} &\colhead{J2000.0} &\colhead{J2000.0}& \colhead{} &
\colhead{$\beta$} & \colhead{$\mu$Jy} & \colhead{$\mu$Jy} &
\colhead{$\mu$Jy} & \colhead{$\mu$Jy} & \colhead{$\mu$Jy} &
\colhead{$\mu$Jy} & \colhead{$\mu$Jy}& \colhead{$\mu$Jy}& \colhead{$\mu$Jy}\\
\colhead{(1)} & \colhead{(2)} & \colhead{(3)} & \colhead{(4)} &
\colhead{(5)} & \colhead{(6)} & \colhead{(7)} & \colhead{(8)} &
\colhead{(9)} & \colhead{(10)}&\colhead{(11)}&\colhead{(12)} &
\colhead{(13)}&\colhead{(14)}
}
\startdata
ID03 &13:24:16.468 &+27:19:07.65 &5.665 &... &0.331$\pm$0.012 &0.353$\pm$0.026   &... &0.291$\pm$0.011 &...&0.286$\pm$0.018&0.334$\pm$0.065&0.254$\pm$0.077&...  \\
ID15 &13:24:23.705 &+27:33:24.82 &5.710 &... &0.492$\pm$0.014 &0.461$\pm$0.025 &... &0.501$\pm$0.014 &...& 0.409$\pm$0.023&0.879$\pm$0.057&0.698$\pm$0.071&...\\
ID20 &13:24:40.527 &+27:13:57.91 &5.724 &... &0.119$\pm$0.012 &0.113$\pm$0.028   &... &...&0.139$\pm$0.012 &0.138$\pm$0.014&0.302$\pm$0.083&...&...  \\
ID23 &13:24:18.450 &+27:16:32.56 &5.922 &... &0.244$\pm$0.011 &0.198$\pm$0.025 &... &... &0.213$\pm$0.010&0.201$\pm$0.013&0.328$\pm$0.048&...&...  \\
ID24 &13:25:19.463 &+27:18:28.51 &6.002 &... &0.244$\pm$0.011 &0.198$\pm$0.025 &... &0.213$\pm$0.010 &...&0.201$\pm$0.013&0.328$\pm$0.048&...&...  \\
ID25 &13:24:26.559 &+27:15:59.72 &6.032 &... &0.291$\pm$0.013 &0.281$\pm$0.026 &... &0.164$\pm$0.018 &...&0.209$\pm$0.021&0.692$\pm$0.045&0.592$\pm$0.071&...  \\
ID27 &13:24:10.766 &+27:19:03.95 &6.040 &... &0.086$\pm$0.012 &...  &...&... &0.098$\pm$0.010&0.069$\pm$0.012&0.191$\pm$0.053&...&...  \\
ID31 &13:23:45.632 &+27:17:00.53 &6.112 &... &0.334$\pm$0.012&...&... &0.116$\pm$0.015&... &0.100$\pm$0.017   &0.344$\pm$0.063 &...&... \\
ID34 &13:23:45.757 &+27:32:51.30 &6.315 &... &0.219$\pm$0.012 &0.350$\pm$0.026  &... &... &0.398$\pm$0.015&0.398$\pm$0.015&1.076$\pm$0.089&...&...  \\
ID35 &13:24:40.643 &+27:36:06.94 &6.332 &... &0.219$\pm$0.012&...&...&... &0.240$\pm$0.013   &0.225$\pm$0.015 &1.854$\pm$0.09&1.159$\pm$0.117&...  \\
ID43 & 13:23:53.054 & +27:16:30.75 & 6.542 & $-3.39\pm0.41$ &
$0.102\pm0.012$ & $0.139\pm0.027$ & $0.152\pm0.018$ & ...&$0.111\pm0.007$ &
 $0.087\pm0.009$ & $0.275\pm0.061$ & ...            & $<$0.263        \\
ID54 &13:24:08.313 &+27:15:43.49 &6.556 &... &0.163$\pm$0.012 &0.219$\pm$0.026   &... &... &0.150$\pm$0.019&0.171$\pm$0.024&0.319$\pm$0.073&0.299$\pm$0.010&...  \\
ID62 &13:23:59.766 &+27:24:55.75 &6.964 &... &0.058$\pm$0.016 &0.394$\pm$0.033&... &0.229$\pm$0.008 &0.247$\pm$0.011&0.242$\pm$0.011&0.421$\pm$0.062&0.310$\pm$0.060&...  \\
\enddata
\tablecomments{The ID numbers are from Table 1 in \citet{Jiang2013}. The $z'$-
and $y'$-band fluxes are from \citet{Jiang2013}, and the remainder from
\citet{Jiang2016,Jiang2020}. The $\beta$ value is also from
\citetalias{Jiang2020}. Some galaxies do not have measurements in some of the
filters, so those columns are left empty.}
\end{deluxetable*}
\end{rotatetable}
\clearpage
\del{
 Removed by s/n<3: ID28 ID30
 Removed by sources near within 1.7 arcsec of IRAC: ID04 ID07 ID36 ID44 ID47 ID49 ID50 ID58 ID61

 Over ID62 and ID25, ID31 not included
 }


\begin{deluxetable*}{ccccccc}[htb!]
\tabletypesize{\footnotesize}
\tablenum{2}
\tablecolumns{1}
\tablecaption{Best-fit parameters and $\chi^2$ values of the 11 galaxies using 
both IRAC bands \label{table5}}
\tablewidth{0pt}
\tablehead{
\colhead{ID} & \colhead{Reduced $\chi^2$} & \colhead{Slope} &
\colhead{Metallicity} & \colhead{Age} & \colhead{Escape fraction} &\colhead{\EBminV}\\ \colhead{} & \colhead{} & \colhead{$\beta$} & 
\colhead{Z} & \colhead{Myr} & \colhead{} &\colhead{mag} \\
\colhead{(1)} & \colhead{(2)} & \colhead{(3)} &
\colhead{(4)} & \colhead{(5)} & \colhead{(6)} &\colhead{(7)}
}
\startdata
ID03  &0.29 &$-$2.41$\pm$0.15 &0.007$\pm$0.012   &210$\pm$180 &0.56$\pm$0.30 &0.1$\pm$0.1  \\
ID15 &0.17 &$-$2.25$\pm$0.14 &0.007$\pm$0.009 &455$\pm$228 &0.47$\pm$0.32 &0.0$\pm$0.10 \\
ID25 &3.26 &$-$2.09$\pm$0.14 &0.005$\pm$0.005   &667$\pm$167 &0.14$\pm$0.20 &0.1$\pm$0.1  \\
ID35 &4.80 &$-$1.21$\pm$0.15 &0.007$\pm$0.008   &567$\pm$177 &0.40$\pm$0.32 &0.6$\pm$0.1 \\
ID54 &0.48 &$-$2.07$\pm$0.22 &0.012$\pm$0.016   &398$\pm$245 &0.49$\pm$0.31 &0.0$\pm$0.1  \\
ID62 &1.11 &$-$2.26$\pm$0.17 &0.013$\pm$0.016   &345$\pm$183 &0.52$\pm$0.31 &0.0$\pm$0.1  \\
\enddata
\centering
\tablecomments{The model parameters along with their Bayesian error ranges of
the 6 galaxies with both IRAC bands included in the fitting.}
\end{deluxetable*}



\begin{deluxetable*}{ccccccc}[htb!]
\tablenum{3}
\tablecolumns{1}
\tablecaption{Best-fit parameters and $\chi^2$ values of the 11 galaxies with
both IRAC bands modeled without IRAC2 at 4.5\micron \label{table6}}
\tablewidth{0pt}
\tablehead{
\colhead{ID} & \colhead{Reduced $\chi^2$} & \colhead{Slope} &
\colhead{Metallicity} & \colhead{Age} & \colhead{Escape fraction} &\colhead{\EBminV}\\ \colhead{} & \colhead{} & \colhead{$\beta$} & 
\colhead{Z} & \colhead{Myr} & \colhead{} &\colhead{mag} \\
\colhead{(1)} & \colhead{(2)} & \colhead{(3)} &
\colhead{(4)} & \colhead{(5)} & \colhead{(6)} &\colhead{(7)}
}
\startdata
ID03  &0.20 &$-$2.35$\pm$0.16&0.007$\pm$0.012 &261$\pm$211 &0.50$\pm$0.31 &0.1$\pm$0.1  \\
ID15 &0.10 &$-$2.16$\pm$0.16&0.006$\pm$0.009 &490$\pm$225 &0.46$\pm$0.32 &0.0$\pm$0.1  \\
ID25 &3.91 &$-$2.12$\pm$0.18&0.006$\pm$0.005 &656$\pm$173 &0.14$\pm$0.19 &0.1$\pm$0.1  \\
ID35 &2.43 &$-$1.08$\pm$0.08& 0.006$\pm$0.005&705$\pm$134 &0.36$\pm$0.30 &0.7$\pm$0.03 \\
ID54 &0.60 &$-$2.08$\pm$0.28&0.012$\pm$0.016 &365$\pm$244 &0.49$\pm$0.32 &0.0$\pm$0.2  \\
ID62 &4.98 &$-$2.22$\pm$0.19& 0.012$\pm$0.016&353$\pm$185  &0.52$\pm$0.31 &0.0$\pm$0.1  \\
\enddata
\centering
\tablecomments{The model parameters along with their Bayesian error ranges of
the same 6 galaxies as in Table~\ref{table5}, but modeled without using the 
IRAC 4.5 \mum\ band.}
\end{deluxetable*}

\begin{deluxetable*}{cccccccc}[h!]
\tablenum{3}
\tablecolumns{1}
\tablecaption{13 SDF galaxy data \label{table4}}
\tablewidth{0pt}
\tablehead{
\colhead{ID}  &\colhead{Reduced $\chi^2$} & \colhead{Slope} &
\colhead{Metallicity} & \colhead{Age} & \colhead{Escape fraction} &\colhead{\EBminV}\\ \colhead{} &\colhead{} & \colhead{$\beta$} & 
\colhead{Z} & \colhead{Myr} & \colhead{} &\colhead{mag} \\
\colhead{(1)} & \colhead{(2)} & \colhead{(3)} &
\colhead{(4)} & \colhead{(5)} & \colhead{(6)} &\colhead{(7)}
}
\startdata
ID03  &0.29 &$-$2.41$\pm$0.15 &0.007$\pm$0.012   &210$\pm$180 &0.56$\pm$0.30 &0.1$\pm$0.10  \\
ID15 &0.17 &$-$2.25$\pm$0.14 &0.007$\pm$0.009 &455$\pm$228 &0.47$\pm$0.32 &0.0$\pm$0.10  \\
ID20 &0.04 &$-$1.80$\pm$0.26 &0.013$\pm$0.017 &345$\pm$239 &0.49$\pm$0.32 &0.4$\pm$0.2  \\
ID23 &0.02 &$-$2.26$\pm$0.15 &0.011$\pm$0.014 &414$\pm$236 &0.41$\pm$0.30 &0.0$\pm$0.1  \\
ID24 &0.03 &$-$2.31$\pm$0.13 &0.010$\pm$0.014 &435$\pm$235 &0.31$\pm$0.26 &0.0$\pm$0.1\\
ID25 &3.26 &$-$2.09$\pm$0.14 &0.005$\pm$0.005   &667$\pm$167 &0.14$\pm$0.20 &0.1$\pm$0.1  \\
ID27&0.48 &$-$2.20$\pm$0.24 &0.007$\pm$0.012   &404$\pm$246 &0.45$\pm$0.31 &0.1$\pm$0.1  \\
ID31 &5.72 &$-$2.59$\pm$0.11 &0.000$\pm$0.001   &381$\pm$285 &0.03$\pm$0.06 &0.0$\pm$0.01  \\
ID34 &0.04 &$-$1.64$\pm$0.20 &0.008$\pm$0.011   &413$\pm$235 &0.47$\pm$0.32 &0.4$\pm$0.1    \\
ID35 &4.80 &$-$1.21$\pm$0.15 &0.007$\pm$0.008   &567$\pm$177 &0.40$\pm$0.32 &0.6$\pm$0.1  \\
ID43  &1.02 &$-$2.28$\pm$0.24 &0.005$\pm$0.007   &443$\pm$246 &0.39$\pm$0.31 &0.0$\pm$0.1  \\
ID54 &0.48 &$-$2.07$\pm$0.22 &0.012$\pm$0.016   &398$\pm$245 &0.49$\pm$0.31 &0.0$\pm$0.1  \\
ID62 &1.11 &$-$2.26$\pm$0.17 &0.013$\pm$0.016   &345$\pm$183 &0.52$\pm$0.31 &0.0$\pm$0.1  \\
\enddata
\centering
\tablecomments{The best-fit parameters and Bayesian values with errors for
implied \fesc\ and UV-slope $\beta$ for the 13 galaxies.}
\end{deluxetable*}


(predominantly OIII and Ha), and that htose are at IRAC wavelengths at these redshifts

\section{SED Modeling}\label{3}

The Lyman-continuum escape fraction may be constrained by the implied presence
of emission lines in the SED of galaxies. A stellar population produces a
certain number of ionizing photons based on its age and stellar mass. If none
of those photons escape (\fesc=0), then the energy will be reprocessed in the
ISM, producing strong emission lines. If the escape fraction is high
(\fesc$\gtrsim$0.3), there will be no associated emission lines
\citep[e.g.,][]{Smith2018, Smith2020, Steidel2018}. CIGALE uses the modeled
stellar age and stellar mass to determine the expected emission line strength,
which may be many Angstroms in equivalent width for these very young stellar
populations, and predicts the associated emission line strengths for a given
\fesc-value. If the near-IR photometry at the emission line wavelengths is
brighter than the CIGALE models predict, then the presence of stronger emission
line is implied, suggesting lower \fesc-values, and vice-versa. Ultimately,
because of uncertainties in the measured stellar population parameters, near-IR
photometry, \textbf{which are at OIII and H$\alpha$ lines for redshifts of our galaxies,} is only weakly dependent on \fesc, resulting in larger errors in the
implied \fesc-values. Deep multi-band photometry including the
\textit{Spitzer} IRAC 3.6--4.5 \micron\ images may thus place some meaningful
constraints on the allowed \fesc-values for these 13 galaxies, enabling us to
explore the average \fesc-value range for galaxies at this redshift, and search
for any trends that may exist between \fesc\ and other galaxy properties. 

Therefore, we used the CIGALE to perform SED modeling for all our SDF galaxies.
We also used the broadband fluxes from \citet{Jiang2016,Jiang2020} for CIGALE
fitting, as described in \S\ \ref{2}. \textbf{CIGALE models fit the Ly$\alpha$ line with a fixed escape fraction directly related to the restframe nebular lines. Despite the uncertainties in \Lya emission modeling, one goal of this paper is to constrain escaping \Lya emission in high redshift galaxies. Therefore, we include the $z’$ photometry which contains important information regarding the strength of escaping \Lya emission relative to other emission lines and keep the Ly$\alpha$ line in our modeling, but increase the uncertainties of WFCAM Z and Y bands by 20 and 10\%, respectively, according to the contribution of \Lya to these bands as listed in \citetalias{Jiang2020}. Our results and conclusions so will be under the assumption that \Lya has been accounted for.}

Following \citetalias{Jiang2020}, we
sampled the model ages between 10 to 900 or 800 Myr using a fixed log
scale with increments of 15, 25, 50, 100, and 200 for redshifts z$>$6.0875 and
z$<$6.0875, respectively, so as to not exceed the age of the Universe at each
redshift. We sampled the escape fraction from 0 to 1 in steps of 0.1. The
metallicity values sampled were 0.0001, 0.0004, 0.004, 0.008, 0.02, and 0.05.
The \EBminV values we sampled were between 0 and 0.7 mag using a
\citet{Calzetti2000} extinction curve. AGN models were not tested since no
significant change was observed in the models by allowing the presence of a
weak AGN in \citetalias{Jiang2020}.  {\bf Lastly, the most relevant part of the modeling is the nebular emission, which is
parametrized by escape fraction and ionization parameter. We varied the ionization parameter between -1.0 and -4.0 in steps of 1 in our modeling.}

For all of our SEDs, we assumed a \citet{Calzetti2000} extinction law. A
discussion of the possible evolution in metallicity and dust extinction is
given by \eg\ \citet{Kim2017}. \citet{Smith2020} and \citet{Oesch2013},
suggesting that SMC extinction curves may provide a better fit to the SED of
star-forming galaxies at $z\simeq2.3-3.5$. However, we found no significant
change in the best-fit SED models when using the SMC extinction curve for our
SDF sample at $z\simeq6$. Fig.~\ref{figure1} presents the SED models of all 13 galaxies by CIGALE.

Dense Basis modeling also used \citet{Calzetti2000} extinction law, and
model parameters such as SFR or \fesc\ were fit to match the observations
instead of testing various values and determining the goodness of fit.
Most galaxy models showed ages greater than 100 Myr and small values of the
extinction, which is consistent with the results of \citet{Jiang2016}, where the
SED models were done with GALEV \citep{kotulla2009}.

Rather than reporting the best-fit values from this analysis (which do
not capture the uncertainties associated with this model fitting well) we
report Bayesian average values and uncertainties associated with each
parameter, as described in \citet[\eg\ ][]{Noll2009}. CIGALE computes such
uncertainties by taking the average value and standard deviation of the fit to
be the average and standard deviation of all models, weighted by the p-value
associated with the model's $\chi^2$ value. These values give a much more
comprehensive picture of the fit than only the best-fit values as it accounts
for the variation in each model rather than considering only one.

One potential concern related to our SED modeling is whether objects with
observations in only one IRAC band would still allow for meaningful
measurements of \fesc\ with our method. To test this, we investigated the \textbf{6
objects from our sample of 13 galaxies with observations in both IRAC bands.} For these objects, we modeled the
SEDs both including and excluding the 4.5 \micron\ band photometry. The results
of this are shown in Tables~\ref{table5} and \ref{table6}, as well as
Fig.~\ref{figure6}. Because of the large uncertainties in the 4.5 \micron\
band photometry, excluding the IRAC2 band photometry --- when available --- in
the modeling did in general not significantly affect the inferred model
parameters. More than half the galaxies overlapped in their implied values for
each parameter using the two sets of models. Thus, we kept the objects without
4.5\micron\ observations in our final sample as long as the remainder data was
of sufficient quality. \del{We apply a similar procedure to test the impact of the
inclusion of K-band data in the 7 galaxies from \citetalias{Jiang2020}, with
the results shown in Tables~\ref{table2} and \ref{table3}. Again, the inclusion
of K-band data does not significantly change the model fits, so the inclusion
of galaxies without K-band data is warranted.}

To further test whether CIGALE could constrain nebular emission without the
IRAC2 band filters, we created simulated models with CIGALE with a known
escape fraction between 0 and 1 for the available filters. The produced flux
values were then rescaled to match our galaxy data, and 100 random variations
for each input escape fraction were produced within the uncertainty of our
photometric uncertainties. The variations were then refit with CIGALE using all
bands except the IRAC2 and K-band filters. The comparison between the original
input escape fraction and refitted output escape fraction are presented in
Fig.~\ref{figure15}, along with a linear regression fit between the input and
output \fesc-values. It can be seen that, while CIGALE does not exactly follow
the input and output escape fractions, excluding the IRAC2 or the K band does
not change the output \fesc\ compared to when all bands are included. The same
procedure was repeated in Fig.~\ref{figure15}b for Dense Basis. Both panels
show similar but not identical relationships between the input and output
\fesc-values with somewhat non-unity slopes. We use the plotted linear
regression fits to map the most likely output \fesc-values onto the input
\fesc-values for both CIGALE and Dense Basis, thus correcting the output from
both modeling methods for this bias. Thus, the emission and \fesc\ may be
constrained by CIGALE or Dense Basis without IRAC2 or K-band data, which is the
case for a part of our sample.

Fig.~\ref{figure7} and \ref{figure8} present the model
parameters corrected for above bias. The two figures also include plots where the
Dense Basis code \citep{Iyer2017} was used instead of CIGALE to determine their
escape fractions to better check for SED fitting bias in CIGALE. The \fesc\
values are corrected for the model bias using the linear regression found in
Fig.~\ref{figure15}. Table \ref{table4} show the results from a Bayesian
treatment of the model parameters of Fig.~\ref{figure1} and list uncertainties
that are needed for the subsequent analysis. 

Fig.~\ref{figure7} shows the distributions of the implied UV-slope $\beta$,
escape fraction \fesc, and extinction values \EBminV. The \fesc\ and $\beta$
values are from the Bayesian values, while \EBminV{} values are the
best-fit $\chi^2$ minimized values. This is because CIGALE uses the $z=0$ slope
of the \EBminV{} vs. $\beta$ relation from the \citet{Calzetti2000} and
\citet{Meurer1999} relations (with an intercept to be determined as a free
parameter from the SED fitting) to estimate \EBminV for a given $\beta$-value,
and so deriving both \EBminV and $\beta$ Bayesian values fully independently is
not possible with this method. The CIGALE models do imply that the intercept
at $z=6$ of the Calzetti/Meurer \EBminV vs. $\beta$ relation at $z=0$ needs to
be $\sim$0.4 bluer in $\beta$ than at z=0 (while adopting the same slope),
which is visible in Fig.~\ref{figure8}a. 

The cumulative Gaussian probability distributions are shown for bluer 
galaxies ($\beta<-2.35$) and redder galaxies ($\beta\ge-2.35$) in the left and
right panels of Fig.~\ref{figure7}, respectively. {\bf Dense
Basis suggest a total range of possible \fesc-values between 0--0.8. Including
the \fesc=0 values at the bottom axes of Fig.~\ref{figure8}b, c, d, and e, the
median escape fraction of CIGALE is about \fesc$\sim$0.55, while median escape
fraction of Dense Basis is about \fesc$\sim$0.35.} Within the currently available
data and photometry, both CIGALE and Dense Basis have significant errors of
$\sim$0.3--0.4 on their \fesc-value estimates for each {\it individual} object.
To within the large individual model fitting errors in \fesc, the two different
model estimates are thus broadly consistent, although not identical. For all
galaxies, the range of median \fesc-values does suggest that the CIGALE and
Dense Basis models imply escape fractions that may be high enough to reionize
the Universe with these UV-bright galaxies \citep{Finkelstein2012,Robertson2015, Naidu2020}.

\del{
Furthermore, {\bf it can be seen that bluer galaxies tend to have lower \fesc\
and \EBminV values. This trend is also seen when the galaxies are not first
divided into two bins of UV-slope $\beta$.}}

Fig.~\ref{figure8} shows comparisons between model parameters for the 13 galaxies and uses their \fesc-values as corrected using the
linear fits in Fig.~\ref{figure15}. Fig.~\ref{figure8}a shows the $z\simeq0$
relation between \EBminV and $\beta$ using the equations that
\citet{Meurer1999} and \cite{Calzetti2000} found at $z\simeq0$. The $z\simeq6$
galaxies of our sample have bluer $\beta$-values than local galaxies at a given
extinction level by about --0.4 in $\beta$-value. This is in line with
expectations, and observations that these galaxies at $z\simeq6$ are bluer
given their particularly younger stellar populations \citep[\eg\
][]{bouwens2014, Labbe2010}. Fig.~\ref{figure8}b shows the relationship
between implied \fesc\ and $\beta$, and Fig.~\ref{figure8}c shows the
relationship between implied \fesc\ and \EBminV. Fig.~\ref{figure8}d and
e show the same as Fig.~\ref{figure8}b and c, but using the Dense Basis
\fesc-values instead. The Dense Base models did not consider $\beta$-values
and assumed \EBminV~$\simeq$0.2 mag, and so only the distribution over their
\fesc-values is shown. No strong trends can be seen between \fesc\ and \EBminV
nor \fesc\ and $\beta$. 

\section{Analysis of the SED Models for the 13 SDF Galaxies at $z\simeq6$}
\label{5}

The sample of 13 SDF galaxies at $z\simeq6$ help to constrain which galaxy population(s) could
have completed and maintained reionization of the Universe at $z\simeq6$.
{\bf Fig.~\ref{figure7} shows cumulative distributions and Gaussians of two
subsamples of the 13 galaxies with the most reliable \fesc{} estimates: bluer
galaxies ($\beta$ $<$ --2.35) and redder galaxies ($\beta$-values $\ge$
--2.35). Fig.~\ref{figure7}a and b show that the bluer subsample has its
distribution peaking around $\beta\simeq$--2.6, while the redder subsample has
its distribution peaking around $\beta\simeq$--2.2. Fig.~\ref{figure7}c and d
show that the bluer subsample has lower implied \EBminV{}-values with a median
\EBminV{}$\simeq$0 mag, while the redder subsample has a higher median near
\EBminV{}$\simeq$0.05 mag. Finally, Fig.~\ref{figure7}e and f show that the
bluer subsample has smaller implied \fesc-values, with a mode and median of
around \fesc$\simeq$0, while the redder subsample has larger implied
\fesc-values of \fesc$\simeq$0.5.}

Fig.~\ref{figure8}b and c shows that our sample galaxies have large uncertainties and
range in the implied \fesc-values ($\sim$0.1--0.35). Thus, there may be some
overlap between the low and high implied \fesc\ galaxies presented in
Fig.~\ref{figure8}. While large photometric and modeling uncertainties
result in large uncertainties in the implied model escape fraction values, 
our analysis does suggest that \fesc\ for $z\sim6$ galaxies is likely less than
$0.8$ in general. \textbf{In fact, our analysis of 13 independent objects with two
independent SED-fitting codes suggests that \fesc-values around 0.5 are
most typical of bright $z\sim6$ galaxies.} This is consistent with the
\fesc-values necessary for bright objects to reionize the Universe by $z\sim6$
\citep{Finkelstein2012,Robertson2015, Naidu2020}.

Fig~\ref{figure8}a indicates that bluer implied $\beta$ corresponds with lower
implied \EBminV values. At given CIGALE-fit \EBminV values, the implied
$\beta$-values at $z\simeq6$ seem to be steeper than those at $z\simeq0$ found
from the Calzetti/Meurer relation \citep{Calzetti2000, Meurer1999} by about
--0.4 in $\beta$. This is expected, and agrees with observations 
\citep[\eg\ ][]{bouwens2014, Labbe2010}, since higher redshift galaxies tend
to have younger stellar populations and lower metallicity and extinction, and
therefore bluer UV-continuum than low redshift galaxies.

Fig.~\ref{figure8}b and~\ref{figure8}c  do not show strong trends between
the parameters. At best, very weak relationships can be seen between redder
$\beta$ or \EBminV with lower \fesc. Fig.~\ref{figure8}d and e using the Dense
Basis \fesc-values do not show any significant trends. This suggests that some
other factors besides nebular emission or dust extinction may contribute to LyC
escaping the galaxies.

One concern was that CIGALE produced larger \fesc\ with redder $\beta$
because it could not distinguish between young galaxies with strong line
emission and mature ones with a larger Balmer break. Misidentifying more mature
galaxies as younger ones could overestimate the line fluxes from the
contribution of the Balmer break, inflating \fesc. To test if this problem
truly existed, a single galaxy was modeled multiple times, but each with flux
values slightly shifted with random values within uncertainty. If CIGALE did
not distinguish the two sets of galaxies, lower \fesc\ should always have
yielded steeper $\beta$-values. However, this was not always the case, as some
models with lower \fesc\ had higher $\beta$ than models with higher \fesc. In
conclusion, there thus seem to be mild correlations at $z\simeq6$ between the
CIGALE model \fesc, $\beta$, and \EBminV{}-values, in the sense that the
implied \fesc\ may trend to higher values for both redder model $\beta$ and for
higher model \EBminV values. 

A range of \fesc-values is possible for any $\beta$ or \EBminV value for both
CIGALE and Dense Basis values. Together with the general lack of clear trends 
between \fesc\ and $\beta$- or \EBminV-values discussed above, the escape
fraction of these galaxies then may depend on some other factors, such as the
geometry or viewing angle, and the porosity of its ISM (see \eg\
\citealt{Smith2018}), that may allow certain lines-of-sight {\it into the
galaxy} to provide higher LyC escape fractions than others \citep{Ma2020}, and
not so much the physical properties of the galaxies. When deeper data over a
wider wavelength range becomes available with JWST, the average opening angle
of the outflow geometry may then be estimated from the median {\bf
\fesc$\sim$0.3--0.55} implied by the CIGALE models in the current work, which
is the \fesc-value needed to complete and maintain the reionization of the
Universe at $z\simeq6$. Cosmological simulations have found that spikes in
\fesc\ come after star-formation episodes, which may indicate channels through
the ISM must first be cleared for LyC to escape \citep[\eg\ ][]{smith2019}.
Because these phenomena that create the ISM openings may differ for individual
galaxies, a large range of \fesc\ could then be produced, as our figures
suggest.

\section{Summary}\label{6}

We extended \textbf{the SED modeling and choose a more robust sample of $z\simeq 6$ {\Lya} emitting galaxies within the
Subaru Deep Field with detailed SED analyses from 7 \cite{Jiang2020} to 13.}
Each of these galaxies had extant ground- and space-based optical--IR
observations of sufficient quality and depth to constrain their SEDs. All SED
fitting was
performed here using the CIGALE package \citep{Boquien2019}. Using the
best-fitting CIGALE SED models, we investigated trends between physical
parameters, such as the fitted UV-slope, $\beta$, the fitted extinction,
\EBminV{}, and the implied escape fractions (\fesc) to infer what factors may
have affected reionization at $z\simeq 6$.

For our sample of \textbf{13} galaxies, the implied ages of the galaxies with uncertainty ranged from \textbf{30 to 800 Myr},
and the implied metallicity from 0.0001 to 0.03, indicating that a wide range
of parameter values is possible for galaxies with blue $\beta$ at $z\simeq6$. 
We found that galaxies at $z\simeq6$ may have a {\it median escape
fraction as implied by CIGALE and Dense Basis}  of {\bf
\fesc$\simeq$0.35--0.55.}\del{, with full a range of 0--0.8. While the current
uncertainties remain large, the likelihood of \fesc-values implied by CIGALE
and Dense Basis do not span the entire range 0--1, and have a median of
\fesc$\sim$0.35--0.55.} The SED fits and \fesc-values may be improved with deeper
observations over a wider wavelength range, such as with the JWST, to improve
the SED models. Furthermore, the
implied $\beta$-values were steeper at given model \EBminV values for
$z\simeq6$ galaxies, when compared to galaxies at $z\simeq0$, as found by
\citet[\eg\ ][]{Calzetti2000} and \citet{Meurer1999}. 

No significant trends were found between \fesc\ and rest of the parameters, given the
significant uncertainties in the data and the modeling. To better constrain
the CIGALE and Dense Basis models, more accurate data and spectroscopic
observations further into the infrared range will be needed. Currently, only
shallow \textit{Spitzer} and Hershel data are available. Furthermore, the
spectra of the galaxies, from which their redshift were derived, does not cover
a wide-enough wavelength range to provide additional constraints to the
available broadband data for better SED fitting. Future deeper spectral
observations, such as with \textit{JWST} NIRSpec at 1--5 $\micron$, would
provide much better constraints to their SED parameters. At $z\simeq6$, the
\textit{JWST/NIRSpec} will be able to observe H$\beta$ and [OIII] emission
lines at 3.0--3.5 \mum, which will provide much better constraints on the
CIGALE models than using broadband fluxes alone. \del{The newly included K-band data
supported the accuracy of the {\bf 7} galaxy models of J20 and improved at
least one model, which demonstrated the need for additional 1.6--5 $\micron$
data points with \textit{JWST} to confirm these results. }

\textbf{We note that besides the high uncertainty in our results, we also assumed that we've correctly modeled the \Lya lines for these galaxies. While we have attempted to correct for this assumption by increasing the uncertainties of filters with \Lya contribution in the modeling, future observations are needed to produce more reliable results.}

Another way to approach this problem will be to use the available spectra to
estimate the physical parameters of these galaxies at $z\simeq6$. The quality
of the spectra will have to be high to make such estimates possible, and to
make a better comparison between the CIGALE models and the available data. With
higher accuracy data and spectroscopy further into the infrared, the
characteristics of galaxies with steep $\beta$ and the escape fractions of
high redshift galaxies can be better characterized to more fully constrain the
sources of the reionization of the Universe at $z\simeq6$.

\acknowledgements{ We thank the two anonymous referees; their reports helped us
clarify and significantly improve our manuscript, and pointed us to the
Anderson-Darling test. We also thank Ian Smail, Pratika Dayal, Eric Gawiser
and Claudia Scarlata for helpful suggestions and several relevant references.
Based on observations made with the NASA/ESA Hubble Space Telescope, obtained
at the Space Telescope Science Institute. Support for HST program GO15137 was
provided by NASA through a grant from the Space Telescope Science Institute,
which is operated by the Association of Universities for Research in Astronomy,
Inc., under NASA contract NAS 5-26555. RAW acknowledges support from NASA
\textit{JWST} Interdisciplinary Scientist grants NAG5-12460, NNX14AN10G and
80NSSC18K0200 from GSFC.}

\facilities{\HST(ACS, WFC3/UVIS, WFC3/IR), Subaru(HSC), \textit{Spitzer}/IRAC,
UKIRT/WFCAM}

\software{astropy \citep{Astropy2018}, CIGALE \citep{Boquien2019} }

\clearpage
\bibliographystyle{aasjournal}

\bibliography{ms5}{}



\begin{figure*}[h!]

\gridline{\fig{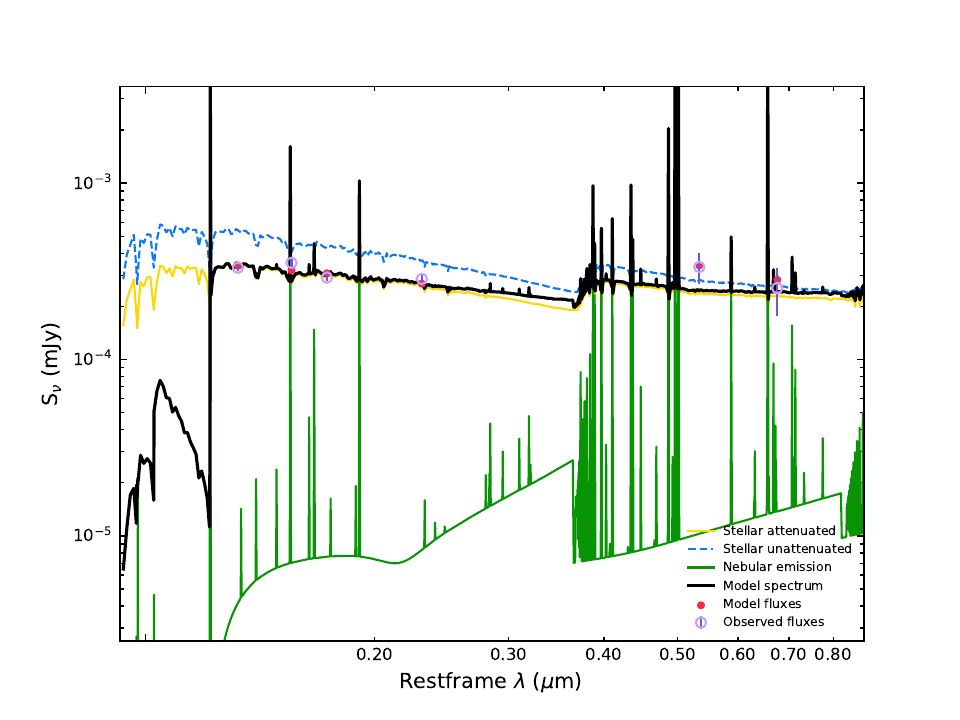}{0.8\textwidth}{(ID03)}
          }
\gridline{\fig{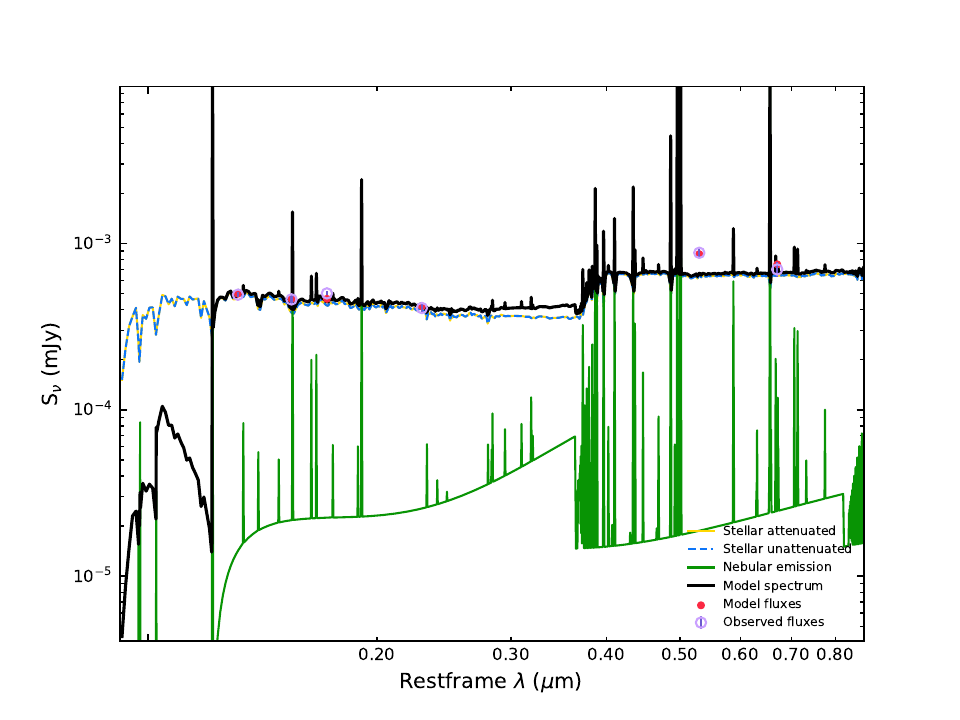}{0.8\textwidth}{(ID15)}
          }
\label{fig:my_label}
\end{figure*}

\begin{figure*}[htb!]

\gridline{\fig{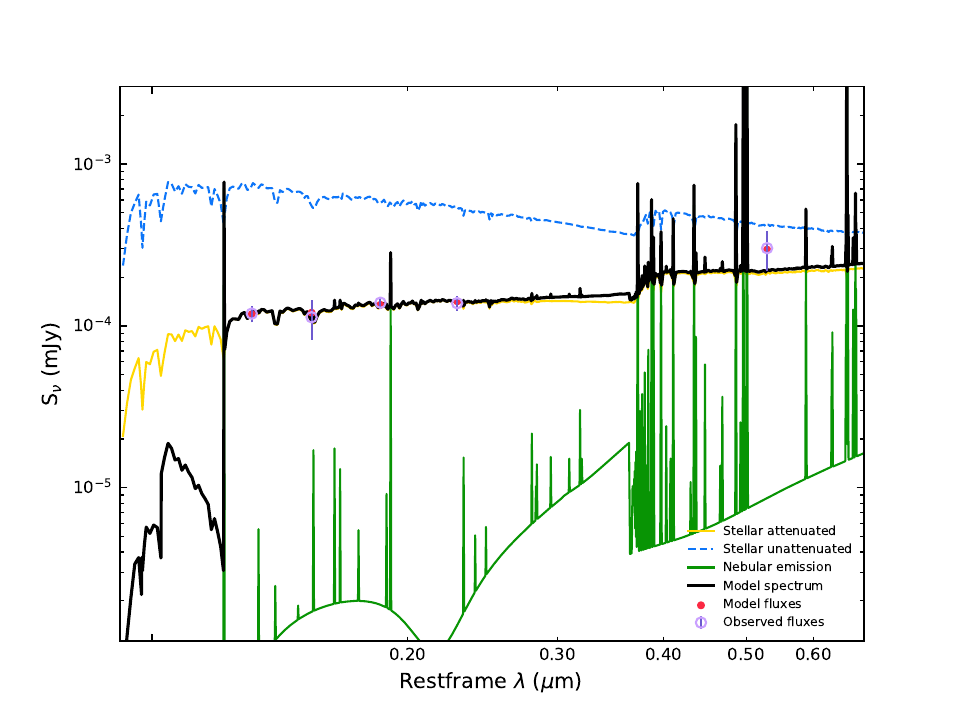}{0.8\textwidth}{(ID20)}
          }
\gridline{
          \fig{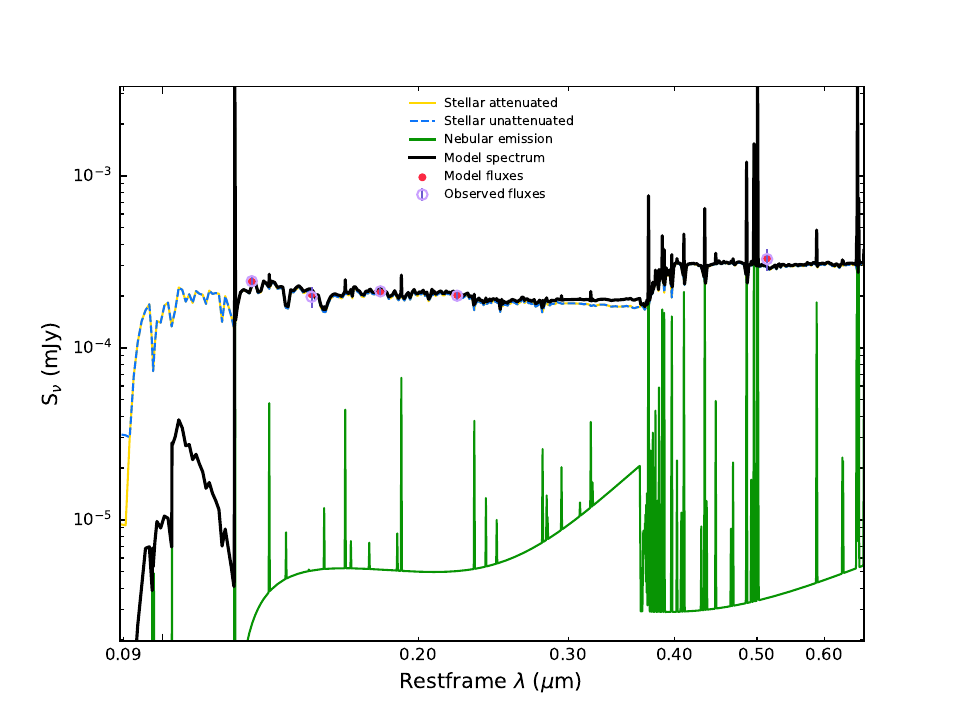}{0.8\textwidth}{(ID23)}
          }
          \end{figure*}
          
\begin{figure*}[htb!]

\gridline{\fig{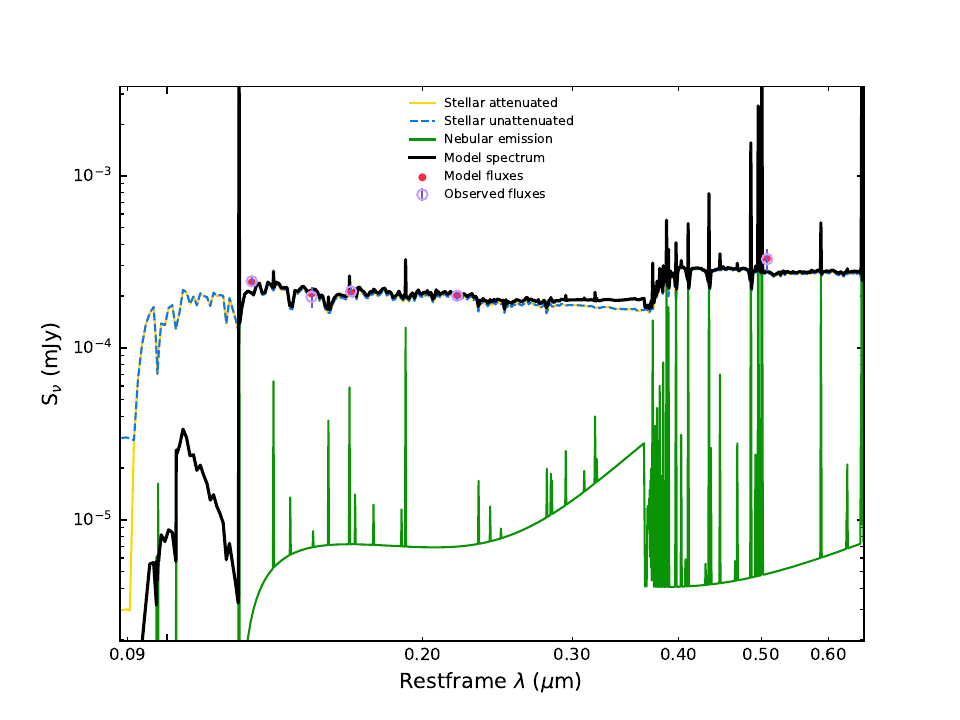}{0.8\textwidth}{(ID24)}
          }
\gridline{\fig{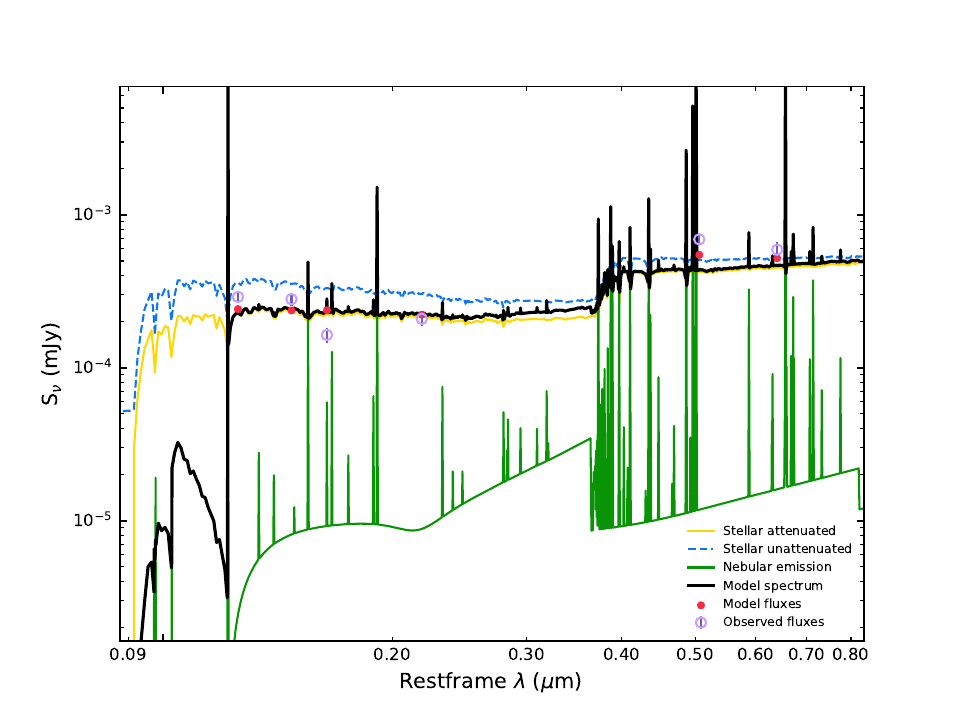}{0.8\textwidth}{(ID25)}}
\end{figure*}
          
\begin{figure*}[htb!]

\gridline{\fig{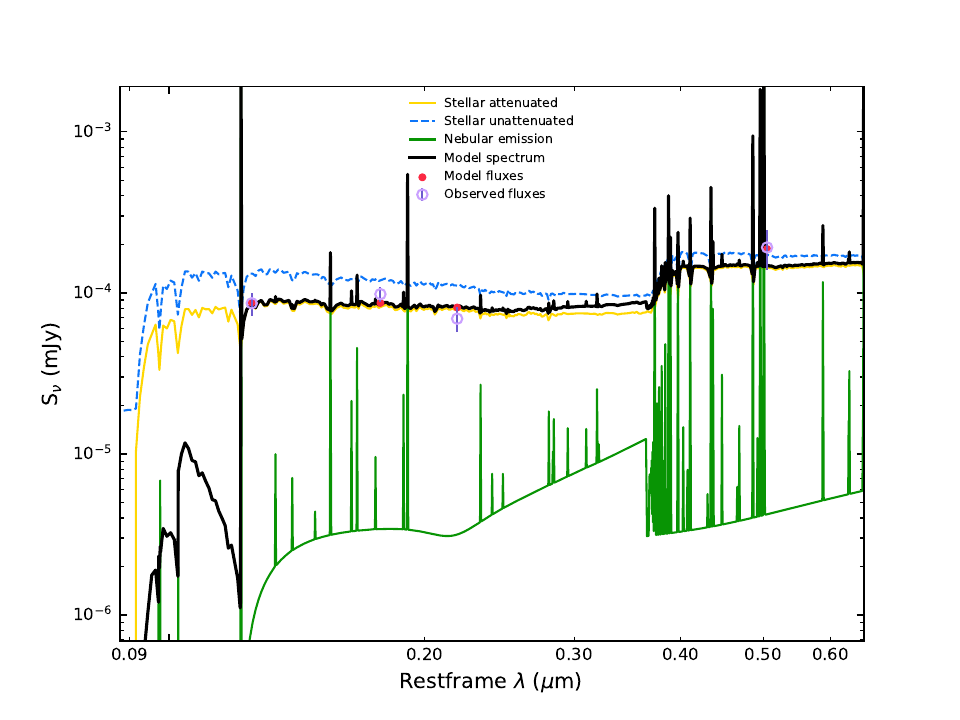}{0.8\textwidth}{(ID27)}
          }
\gridline{\fig{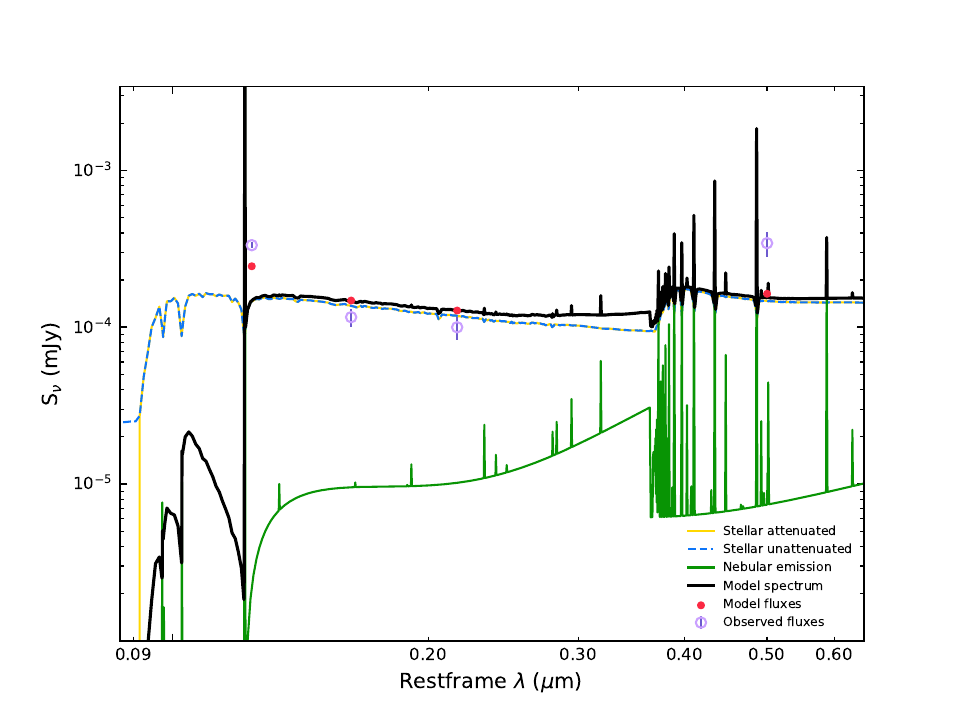}{0.8\textwidth}{(ID31)}
          }
\end{figure*}

\begin{figure*}[htb!]

\gridline{\fig{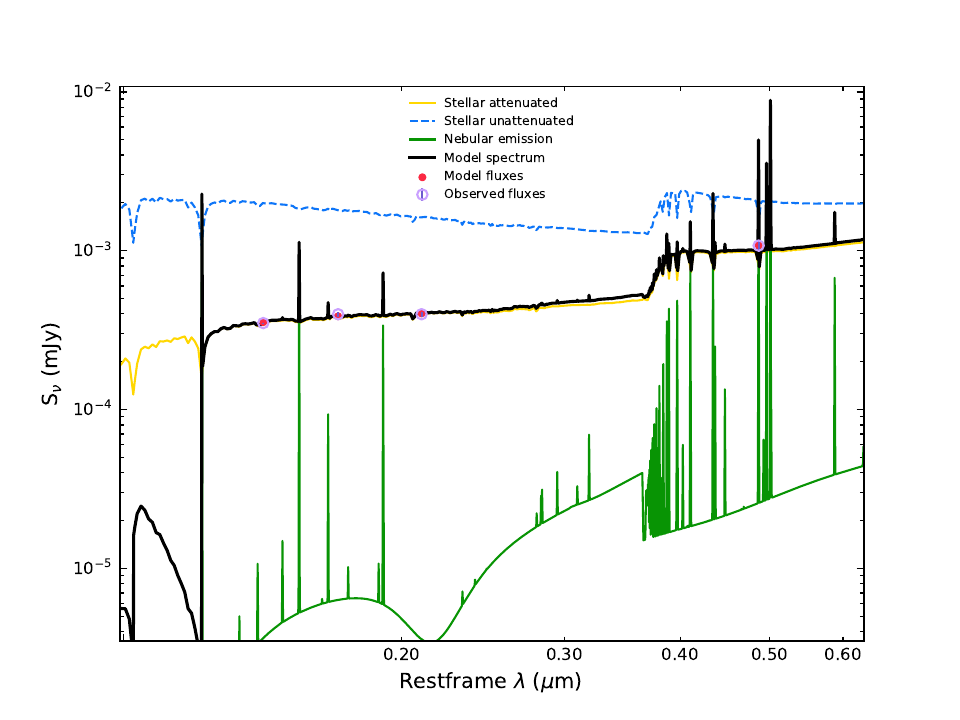}{0.8\textwidth}{(ID34)}
          }
\gridline{\fig{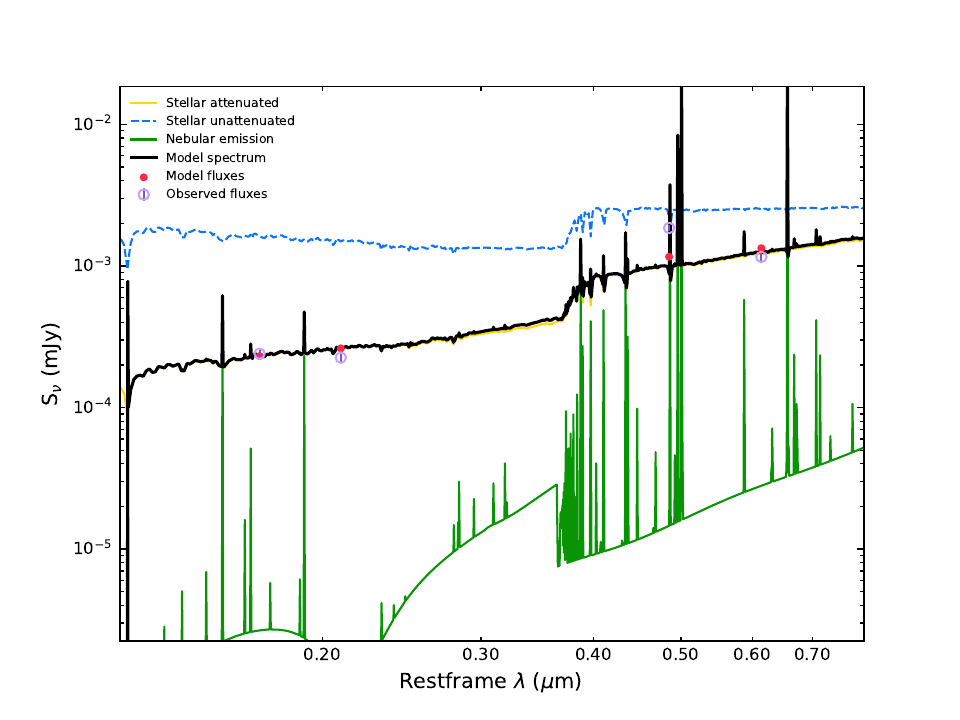}{0.8\textwidth}{(ID35)}}
\end{figure*}

\begin{figure*}
\gridline{
          \fig{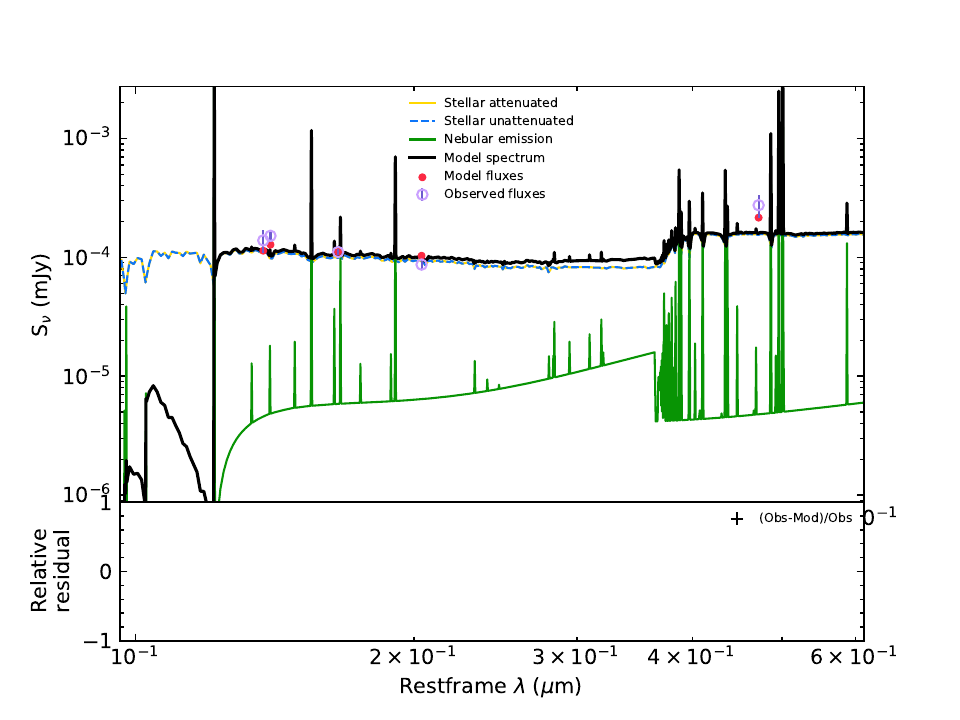}{0.8\textwidth}{(ID43)}
          }

\gridline{\fig{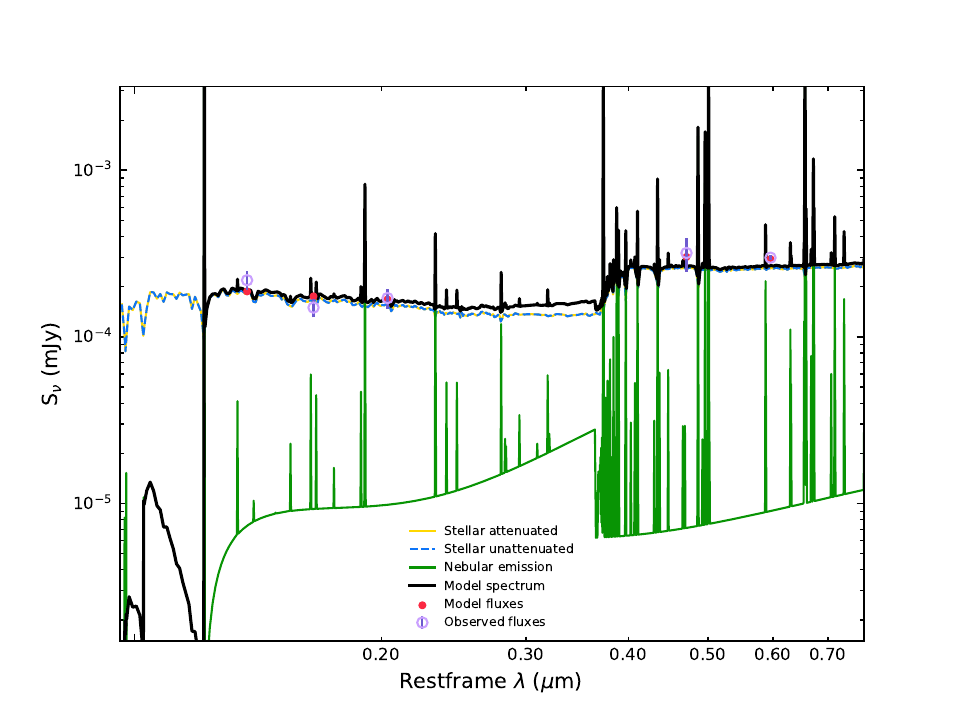}{0.8\textwidth}{(ID54)}
          }
          \end{figure*}
          
\begin{figure*}
\gridline{\fig{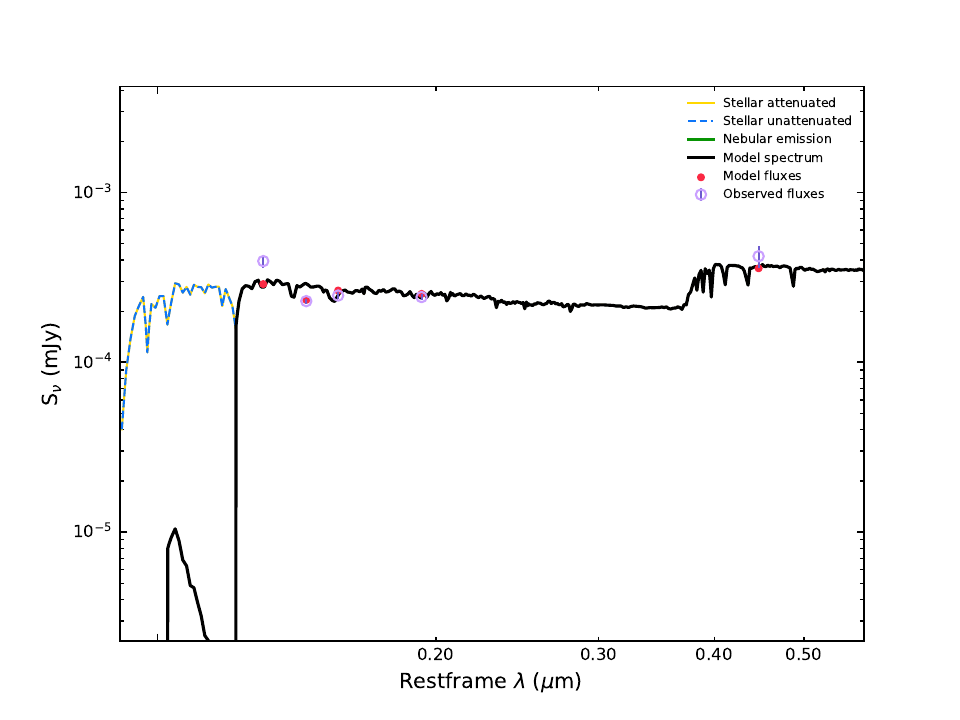}{0.8\textwidth}{(ID62)}}
\end{figure*}

\begin{figure*}[htb!]
\figsetstart
\figsetnum{1}
\figsettitle{SED Models of the 13 galaxies}

\figsetgrpstart
\figsetgrpnum{figurenumber.9}
\figsetgrptitle{ID03}
\figsetplot{ID03_best_model.pdf}
\figsetgrpend

\figsetgrpstart
\figsetgrpnum{figurenumber.13}
\figsetgrptitle{ID15}
\figsetplot{ID15_best_model.pdf}
\figsetgrpend

\figsetgrpstart
\figsetgrpnum{figurenumber.15}
\figsetgrptitle{ID20}
\figsetplot{ID20_best_model.pdf}
\figsetgrpend

\figsetgrpstart
\figsetgrpnum{figurenumber.18}
\figsetgrptitle{ID23}
\figsetplot{ID23_best_model.pdf}
\figsetgrpend

\figsetgrpstart
\figsetgrpnum{figurenumber.19}
\figsetgrptitle{ID24}
\figsetplot{ID24_best_model.pdf}
\figsetgrpend

\figsetgrpstart
\figsetgrpnum{figurenumber.20}
\figsetgrptitle{ID25}
\figsetplot{ID25_best_model.pdf}
\figsetgrpend

\figsetgrpstart
\figsetgrpnum{figurenumber.21}
\figsetgrptitle{ID27}
\figsetplot{ID27_best_model.pdf}
\figsetgrpend

\figsetgrpstart
\figsetgrpnum{figurenumber.23}
\figsetgrptitle{ID31}
\figsetplot{ID31_best_model.pdf}
\figsetgrpend

\figsetgrpstart
\figsetgrpnum{figurenumber.25}
\figsetgrptitle{ID34}
\figsetplot{ID34_best_model.pdf}
\figsetgrpend

\figsetgrpstart
\figsetgrpnum{figurenumber.26}
\figsetgrptitle{ID35}
\figsetplot{ID35_best_model.pdf}
\figsetgrpend

\figsetgrpstart
\figsetgrpnum{figurenumber.4}
\figsetgrptitle{ID43}
\figsetplot{ID43_best_model_k.pdf}
\figsetgrpend

\figsetgrpstart
\figsetgrpnum{figurenumber.39}
\figsetgrptitle{ID54}
\figsetplot{ID54_best_model.pdf}
\figsetgrpend

\figsetgrpstart
\figsetgrpnum{figurenumber.41}
\figsetgrptitle{ID62}
\figsetplot{ID62_best_model.pdf}
\figsetgrpend

\figsetend
\caption{The best-fit SED models of the 13 galaxies created using CIGALE
\citep{Boquien2019}. The blue boxes are the observed fluxes while the red
circles are the model fluxes. The uncertainties plotted are the $3\sigma$ uncertainties. In general, the models are able to fit the
observations, providing meaningful constraints on the extinction, UV-slope
$\beta$, and escape fraction. Furthermore, the models produce redder implied
slopes than \citetalias{Jiang2020}.
\label{figure1}}
\end{figure*}



\begin{figure*}[h!]

\gridline{
 \fig{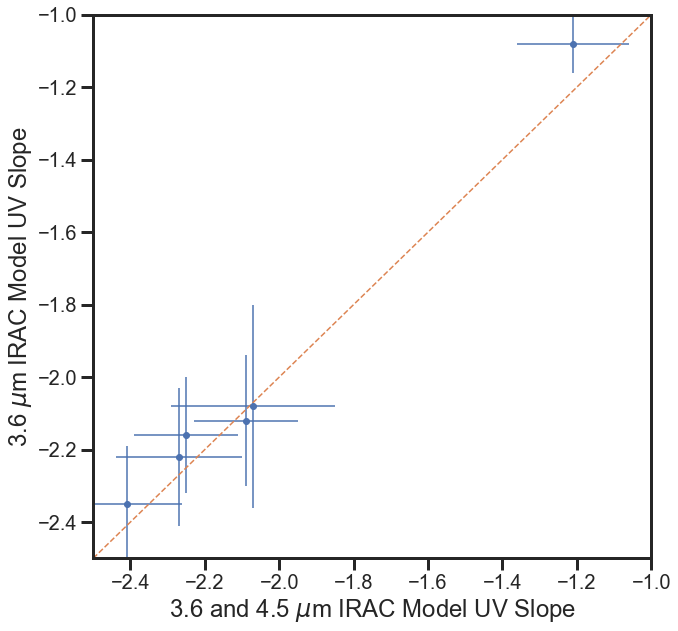}{0.33\textwidth}{(d)}
 \fig{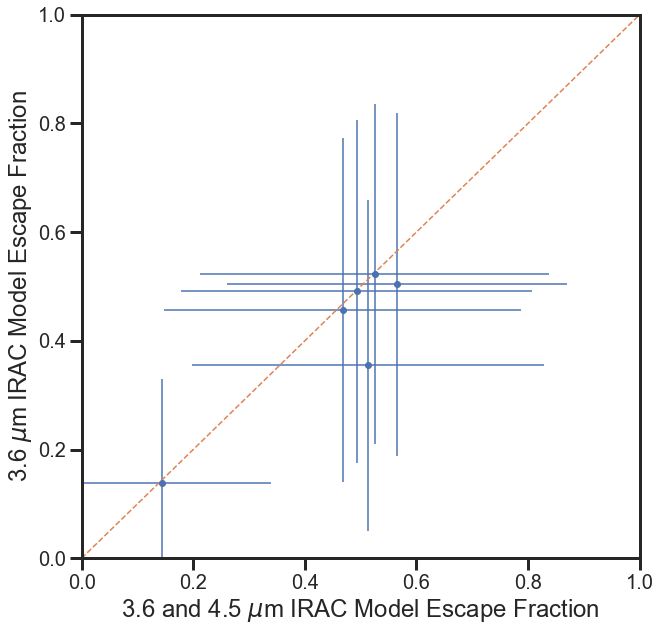}{0.33\textwidth}{(e)}
 \fig{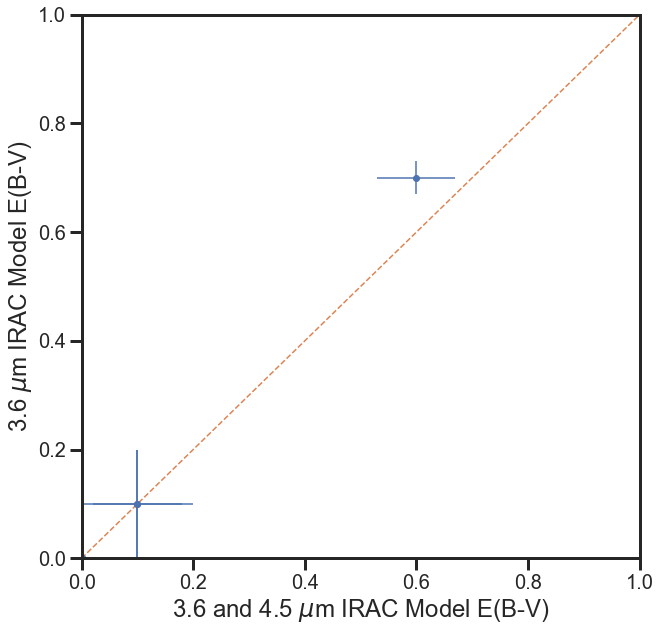}{0.33\textwidth}{(f)}
}
\caption{Comparison of $\beta$, \fesc, and \EBminV of the models fit when both
3.6 and 4.5\micron\ IRAC filters are included and when only 3.6\micron\ is
included, using the data of Tables~\ref{table5} -- \ref{table6}. Only minor
differences can be seen in the two sets of models, showing that the addition of
the much noisier IRAC2 4.5 \micron\ data points does not affect the modeling
significantly. 
\label{figure6}}
\end{figure*}



\begin{figure*}[h!]
\gridline{
 \fig{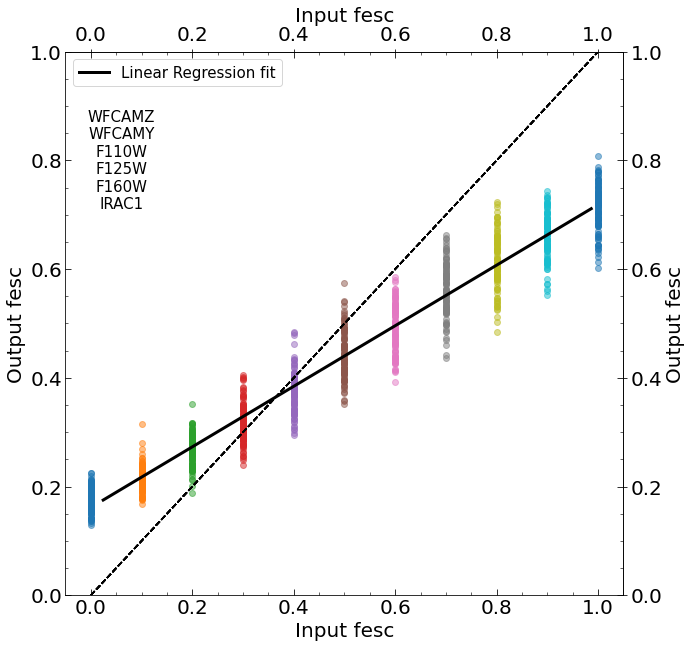}{0.5\textwidth}{(a)}
 \fig{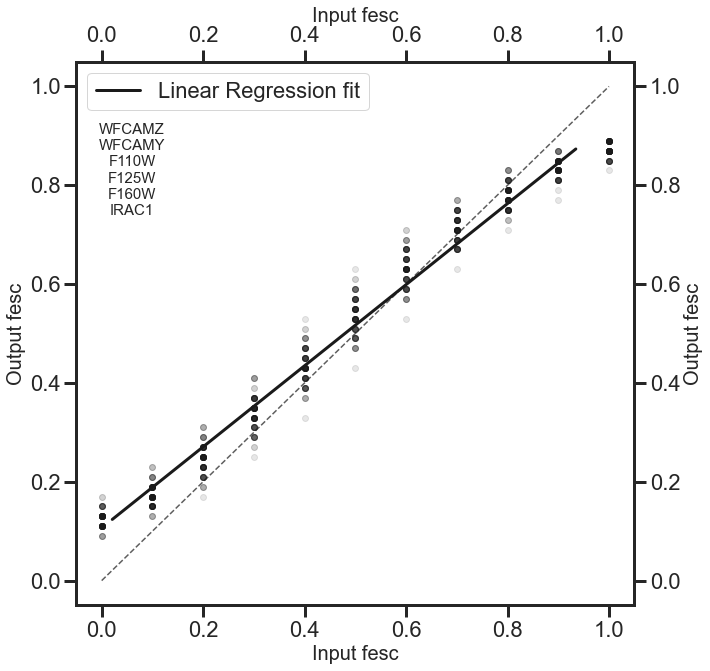}{0.5\textwidth}{(b)}
}
\caption{
 (a) [Left panel]:\ Output escape fraction using the CIGALE models
against a range of input escape fractions. This was done by refitting
100 random variations of the original models with a fixed escape fraction created by CIGALE using the photometry in all
filters plus its errors, but excluding the IRAC2 and K-band filters, which had
the lowest S/N-ratio and did not constrain the SED fits well. The linear regression fit has an equation of \fesc $_{,in}$ = 1.79\fesc $_{,out}-$0.29 
{\bf (b) [Right panel]:}\ Same as (a) for Dense Basis. Both panels show similar
but not identical relationships between the input and output \fesc-values. We 
use the plotted linear regression fits to map the most likely output
\fesc-values onto the input \fesc-values for both CIGALE and Dense Basis, thus
correcting both modeling methods for this bias. The linear regression fit has an equation of \textbf{\fesc $_{in}$ = 1.22\fesc $_{out}-$0.13}}
\label{figure15}
\end{figure*}



\begin{figure*}[h!]
\vspace{-0.30cm}
\gridline{ 
 \fig{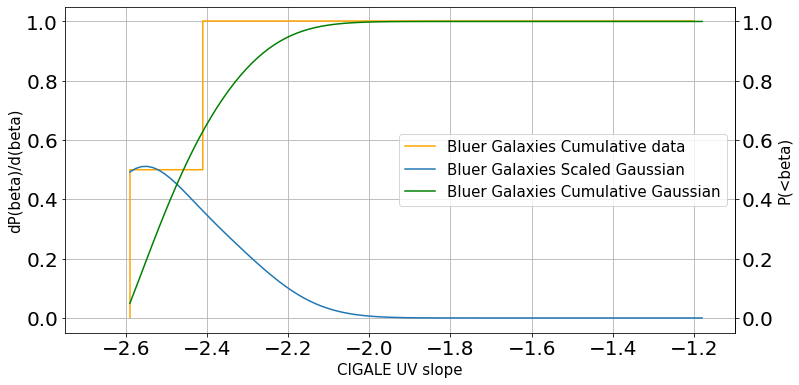}{0.5\textwidth}{(a)}
 \fig{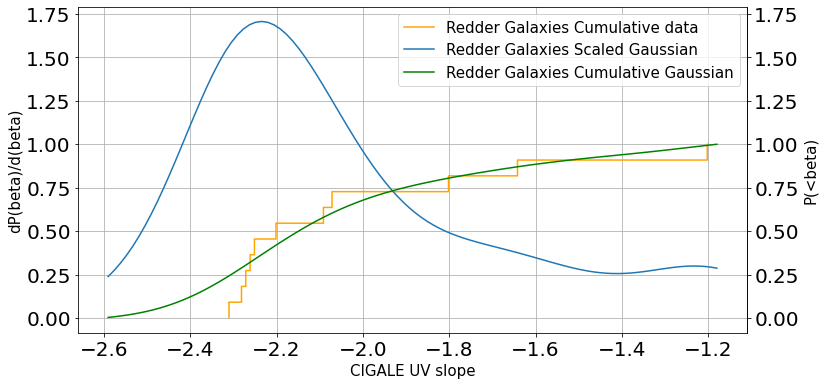}{0.5\textwidth}{(b)}
}
\vspace{-0.30cm}
\gridline{
 \fig{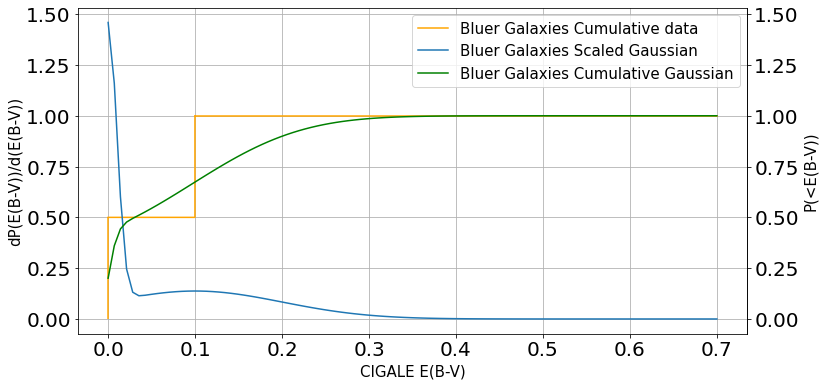}{0.5\textwidth}{(c)}
 \fig{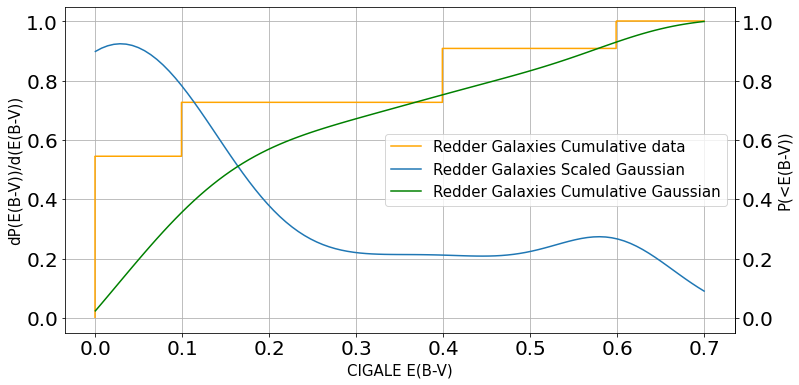}{0.5\textwidth}{(d)}
}
\vspace{-0.30cm}
\gridline{
 \fig{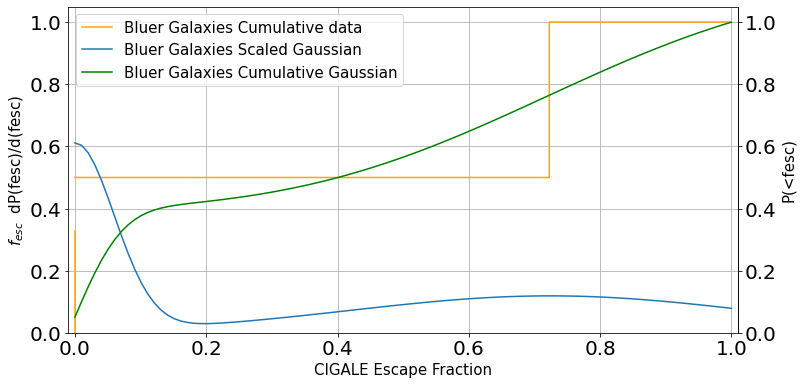}{0.5\textwidth}{(e)}
 \fig{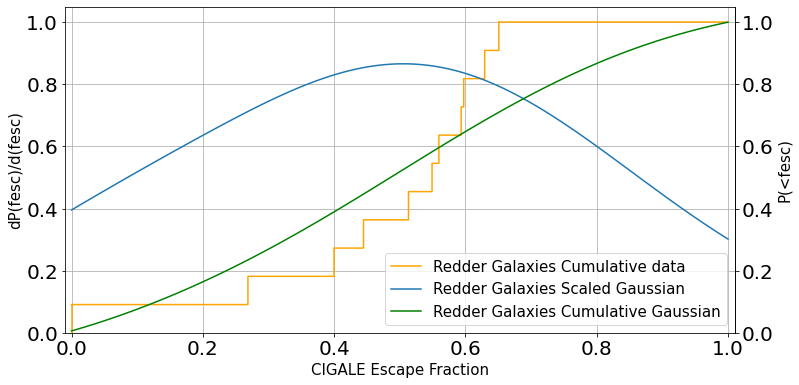}{0.5\textwidth}{(f)}
}
\vspace{-0.30cm}
\gridline{
 \fig{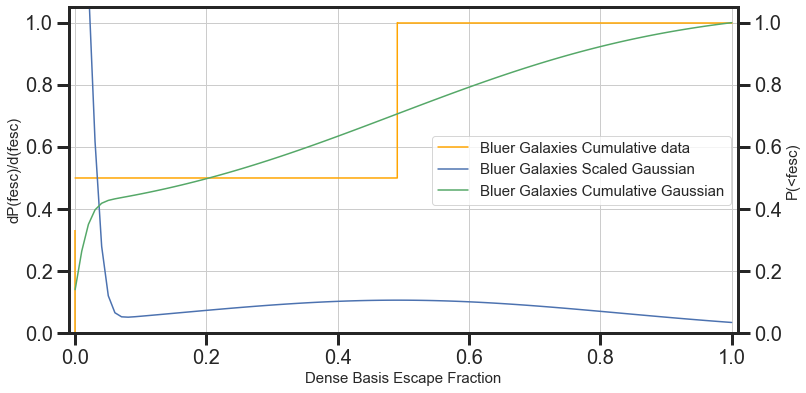}{0.5\textwidth}{(g)}
 \fig{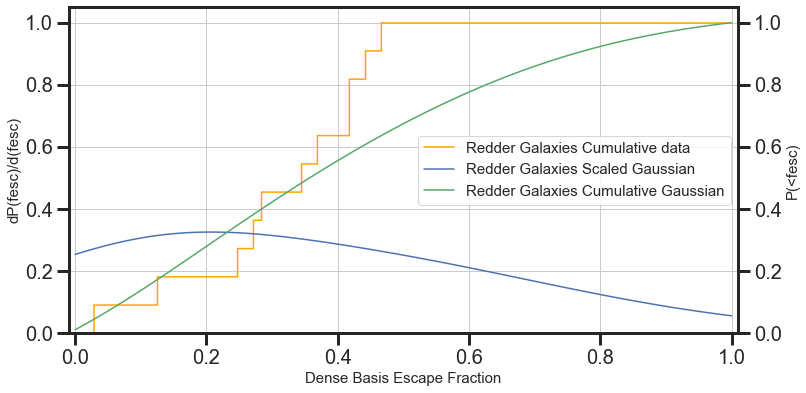}{0.5\textwidth}{(h)}
}
\vspace{-0.30cm}
\caption{Distributions of the implied Bayesian UV-slope and escape
fraction, and best-fit \EBminV lines of the 13 galaxies. The sample is divided into bluer
galaxies with model $\beta<-2.35$ and redder galaxies with model
$\beta\ge-2.35$. Panels (g) and (h) show the distribution of the Dense Basis 
\fesc-values. The orange lines show the cumulative distributions of the
individual adopted values with the probability shown on the right axis. The
blue and green lines respectively show the scaled stacked model likelihoods and
stacked cumulative distributions with probability for the latter on the left
axis, assuming that the likelihood of the model parameter is normally
distributed with the given mean and standard deviation.  The CIGALE
plots show that bluer galaxies --- as selected --- are more likely to have lower
values of \fesc\ and lower \EBminV{}. (The Dense Base models did not consider
$\beta$-values and assumed \EBminV$\simeq$0.2 mag, and so only the distribution
over their \fesc-values is shown).}
\label{figure7}
\end{figure*}



\vspace*{-1.00cm}
\begin{figure*}[h!]
\n\cl{
 \includegraphics[width=0.400\txw]{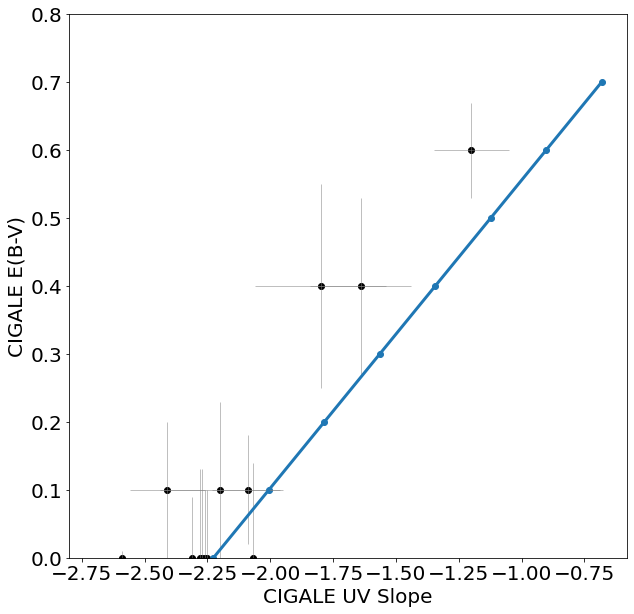}
}
\n\cl{
 \includegraphics[width=0.400\txw]{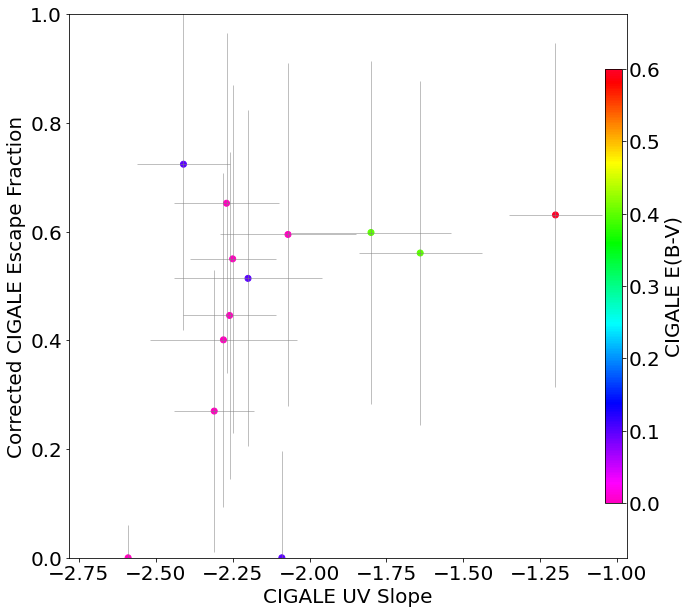}\ \ 
 \includegraphics[width=0.410\txw]{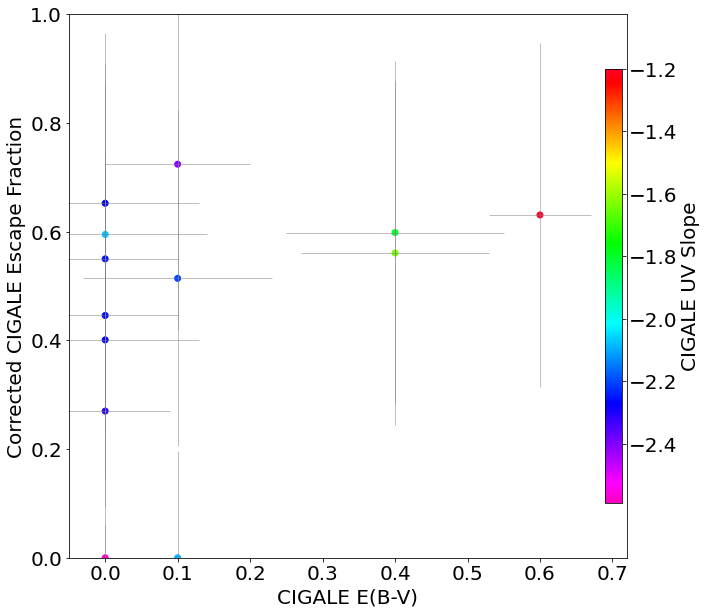}
}
\n\cl{
 \includegraphics[width=0.400\txw]{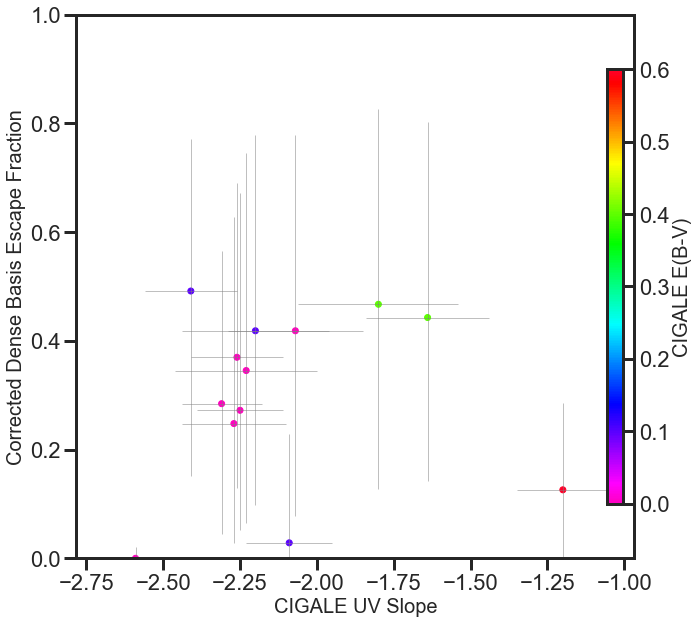}\ \ 
 \includegraphics[width=0.410\txw]{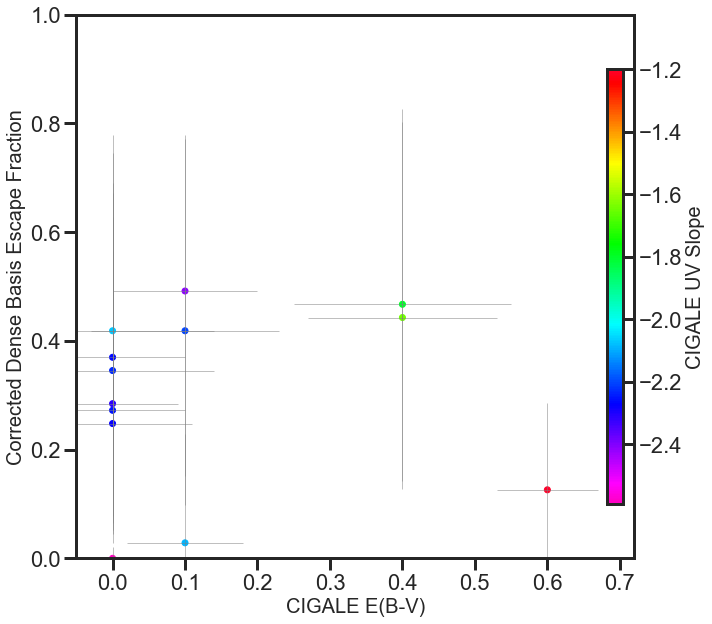}
}
\caption{
 Distribution of the Bayesian $\beta$ and \fesc, and best-fit \EBminV
for the \textbf{13} galaxies. 
{\bf (a) [Top panel]:}\ Best fit CIGALE \EBminV vs. $\beta$, which tends to be
steeper (bluer) for lower implied \EBminV values. The local relation between
\EBminV and $\beta$ from \citet{Calzetti2000, Meurer1999} is plotted in blue,
showing shallower $\beta$-values at $z\simeq0$ compared to our SDF sample at
$z\simeq6$. 
{\bf (b),(c) [Middle panels]:}\ CIGALE \fesc-values as derived from the linear
fit of Fig.~\ref{figure15}. 
{\bf (d),(e) [Bottom panels]:}\ Dense Basis \fesc-values as derived from the
linear fit of Fig.~\ref{figure15}. These figures suggest that the possible
range of \fesc\ is between 0 to 0.8 with a median of \textbf{0.35--0.55}. No significant
trends between \fesc\ and UV-slope or \EBminV are present. If the nebular
emission or the dust extinction has no discernible effect on \fesc, other
factors such as the distribution of holes vacated in the ISM by supernovae
and/or weak AGN somewhat later in a galaxy's evolutionary stage may be needed
to let LyC radiation escape.}

\label{figure8}
\end{figure*}


\end{document}